\documentclass[journal]{IEEEtran}

\usepackage[T1]{fontenc} 
\usepackage[utf8]{inputenc}
\usepackage{pgfplots}
\usepackage{tikz}
\usepackage{pgfplotstable}
\usepgfplotslibrary{groupplots}
\usepackage{siunitx}
\usepackage{amsmath}

\ifCLASSINFOpdf
\graphicspath{{plots/}}
\else
\fi
\ifCLASSOPTIONcompsoc
 \usepackage[caption=false,font=normalsize,labelfont=sf,textfont=sf]{subfig}
\else
 \usepackage[caption=false,font=footnotesize]{subfig}
\fi
\usepackage{url}

\usepackage{amsfonts}
\usepackage{amsmath}
\usepackage[disable]{todonotes}  
\usepackage{bbold}
\usepackage[normalem]{ulem}
\usepackage{threeparttable}  
\usepackage[export]{adjustbox}  
\usepackage{hyperref}

\definecolor{babyblueeyes}{rgb}{0.63, 0.79, 0.95}
\definecolor{bittersweet}{rgb}{1.0, 0.44, 0.37}
\definecolor{caribbeangreen}{rgb}{0.0, 0.8, 0.6}
\definecolor{celadon}{rgb}{0.67, 0.88, 0.69}
\definecolor{babypink}{rgb}{0.96, 0.76, 0.76}

\newcommand{\jntodo}[1]{\todo[inline, color=caribbeangreen]{@JN #1}}
\newcommand{\jncomment}[1]{\todo[inline, color=celadon]{JN: #1}}

\begin{document}

\title{Generalised Score Distribution:\\A Two-Parameter Discrete Distribution
Accurately Describing Responses from Quality of Experience Subjective
Experiments}

\author{Jakub~Nawała, Lucjan~Janowski, Bogdan~Ćmiel, Krzysztof~Rusek, Pablo~Pérez
	\thanks{This work is licensed under a \href{http://creativecommons.org/licenses/by/4.0/}{Creative Commons Attribution 4.0 International License}.}%
	\thanks{The research leading to these results has received funding from the
	Norwegian Financial Mechanism 2014-2021 under project 2019/34/H/ST6/00599.
	Furthermore, the work was supported by the PL-Grid Infrastructure.}%
	\thanks{J.~Nawała, L.~Janowski, and K.~Rusek are with the Institute of Telecommunictions, AGH University of Science and Technology, Kraków, Poland. (e-mail: jakub.nawala@agh.edu.pl)}
	\thanks{B.~Ćmiel is with the Department of Mathematical Analysis,
	Computational Mathematics and Probability Methods,
	AGH University of Science and Technology, Kraków, Poland.}%
	\thanks{P.~Pérez is with Applications \& Platforms Software Systems, Nokia Bell Labs, Madrid, Spain.}}
	
\markboth{Submitted to IEEE Transactions on Multimedia}%
{Nawała \MakeLowercase{\textit{et al.}}: Generalised Score Distribution}
%

\maketitle

\begin{abstract}
Subjective responses from Multimedia Quality Assessment (MQA)
experiments are conventionally analysed with methods not suitable
for the data type these responses represent. Furthermore, obtaining
subjective responses is resource intensive. A method allowing reuse
of existing responses would be thus beneficial.
Applying improper data analysis methods leads to difficult to
interpret results. This encourages drawing erroneous conclusions.
Building upon existing subjective responses is resource friendly and
helps develop machine learning (ML) based visual quality predictors.
We show that using a discrete model for analysis of responses from
MQA subjective experiments is feasible. We indicate that our
proposed Generalised Score Distribution (GSD) properly describes
response distributions observed in typical MQA experiments. We
highlight interpretability of GSD parameters and indicate that the
GSD outperforms the approach based on sample empirical distribution
when it comes to bootstrapping. We evidence that the GSD outcompetes
the state-of-the-art model both in terms of goodness-of-fit and
bootstrapping capabilities.
To do all of that we analyse more than one million subjective
responses from more than 30 subjective experiments. Furthermore,
we make the code implementing the GSD model and related analyses
available through our GitHub repository:
\url{https://github.com/Qub3k/subjective-exp-consistency-check}.
\end{abstract}


\section{Introduction}
\label{sec:introduction}
%
%
%
%
%
\IEEEPARstart{T}{here} are phenomena that require gathering opinions from
a panel of people. One significant example here is the notion of Quality of
Experience (QoE). Contrary to the Quality of Service (QoS), the QoE also
depends on how a user of a system perceives its performance (with the word
\textit{perceives} being the most important here). Although technical factors
do influence the QoE, ultimately, it is a subjective opinion of a user that
represents the most direct indication of the QoE.
(See Sec. 2.2.2 of~\cite{QoEBook} for a formal definition of the QoE.)

Multimedia Quality Assessment (MQA) is a sub-field of the QoE related research activities.
It focuses on understanding how people perceive quality of multimedia content,
effects of its processing and performance of multimedia services. It is a
common and recommended~\cite{P913_2021}\cite{BT500} practice to organise
experiments in which
a panel of observers provides its opinion on quality of multimedia materials
presented. We refer to such experiments as \textit{subjective experiments}
and to opinions provided by the panel of observers as \textit{subjective
responses}. Importantly, we narrow our discussion down to subjective experiments
in which participants judge technical reproduction quality of stimuli
presented only. In other words, we do not take into account subjective
experiments, where observers voice their opinion regarding content of
stimuli (e.g., plot of a story or artistic properties of an image).

Lack of access to ground truth information is an inherent feature of
subjective experiments. Differently put, we observe subjective responses, but
have no way of directly measuring quality of a given stimulus.
One solution to this problem is
gathering a large number of responses per stimulus. This way, we make sure
any summary statistic we use to estimate stimulus quality is well reflecting
population level opinion. More and more researchers follow this intuition
and switch from small scale controlled experiments to large scale
crowdsourcing experiments~\cite{Ying2021}.
Unfortunately, switching to crowdsourcing experiments usually corresponds
to less precise measurements. On the other hand, organising large scale
controlled subjective
experiments is money- and time-intensive. For these reasons we want to
learn as much as possible from limited information controlled subjective
experiments provide.

To fully use information controlled subjective experiments provide
we cannot rely on summary statistics only (from which the Mean Opinion Score
or MOS is the most popular~\cite{P913_2021}\cite{BT500}). Instead, we need to construct
models that try to capture the underlying, unobservable structure of subjective
responses and how this structure maps to quality. To construct such models
we use various assumptions based on domain knowledge and experiences from
previous subjective experiments.

There are better and worse models. Likewise, there are tools to assess how
well a model performs. We claim that using models reflecting data type
subjective responses represent is a better approach than assuming that
continuous models can be applied to discrete data. For one thing, models
reflecting underlying data type generate interpretable results.
Increased interpretability makes it easier to understand the result and
thus protects against ill posed conclusions.

\subsection{Problem Statement and Contributions}
\label{ssec:problem_statement_and_contributions}
Subjective responses from Multimedia Quality Assessment (MQA)
experiments are conventionally analysed with methods not suitable for data type
these responses represent. In particular, continuous models are used even though
subjective responses are discrete in most cases~\cite{Li2017}\cite{Li2020EI}.
Furthermore, obtaining
subjective responses is money- and time-consuming. A method allowing to reuse and
build upon existing subjective responses would be thus beneficial.

Applying improper data analysis methods may lead to results that are difficult to
interpret. This may result in erroneous conclusions. Liddell and Kruschke
provide a convincing overview of
mistakes that arise when analysing data using an improper
model~\cite{Liddell2018}. One of our goals is to protect researchers
analysing responses from MQA experiments against mistakes Liddell and Kruschke
mention.

If it comes to building upon existing subjective responses, the approach is
especially important if used to
generate large samples from small real-life samples. This procedure is also
referred to as \textit{bootstrapping}. Properly applied bootstrapping
gives a chance
of producing sample sizes sufficient for developing machine learning (ML) visual
quality predictors. Naturally, reusing existing subjective responses is also
resource friendly.

We show that using a discrete model for data analysis of responses from MQA
subjective experiments is feasible. We also present benefits stemming from this
approach. Specifically, we indicate that our proposed Generalised Score
Distribution (GSD) properly describes response distributions observed in typical
MQA experiments. We also highlight interpretability of GSD parameters. This GSD
feature makes it possible to easily describe and intuitively understand
non-trivial dependencies between various response distributions. At last, we
point out that the GSD outperforms the traditional approach based on a sample
empirical distribution when it comes to bootstrapping.

Being more suitable for bootstrapping than sample empirical distribution the GSD
can generate a large data set of responses by taking advantage of only a small
data set of real-life responses. In turn, large data sets generated this way may
give a chance to create next generation ML based perceptual visual quality
predictors. This is because ML based solutions are data hungry by nature, and
typical MQA experiments are capable of gathering only few dozens of responses per
stimulus. Moreover, knowing a correct data model (which we show the GSD is for
typical MQA experiments) allows proposing a parametric hypothesis testing
framework. Using such a framework results in higher power (when compared to
conventionally used here non-parametric methods) and thus allows to reduce costs
related to organising
subjective experiments. This is because more powerful statistical methods allow
to detect smaller effect sizes, while keeping the sample size constant. Finally,
interpretability of GSD parameters makes it easier to summarise subjective
responses and perform non-misleading intuitive inferences based on this summary.

With this work we put forward the following contributions:
\begin{itemize}
    \item We evidence that analysing subjective responses from MQA experiments
    with a discrete model (specifically, the GSD) is feasible and brings easy to
    interpret results.
    \item We indicate that the GSD well describes responses from typical MQA
    subjective experiments.
    \item We show that the GSD outperforms empirical distribution when it comes
    to bootstrapping for responses from MQA subjective experiments.
    \item At last, we demonstrate that the GSD outperforms the
    state-of-the-art model when it comes to bootstrapping and
    goodness-of-fit testing.
\end{itemize}

The main idea of the paper is that we want to convince the MQA research community
that using the GSD model to analyse subjective responses is better than
following current recommendations and practices put forward in the literature.
Specifically, we want to show that the GSD outperforms non-disrete models
used in the literature and, when it comes to bootstrapping, that the GSD also
performs better than the standard approach based on a sample empirical
distribution. To make it easier for others to use our work, we invite
everyone to visit our GitHub repository
(\url{https://github.com/Qub3k/subjective-exp-consistency-check}). There,
we provide a code allowing to analyse subjective responses with the use
of the GSD model and to reproduce a significant part of results
presented in this paper.

\subsection{Related Works}
There is a trend in the MQA community to favour response distribution analysis
over relying on summary statistics (e.g., the MOS) only.  An important recent
contribution in this topic is the work by Seufert~\cite{Seufert2021}. There, he
highlights fundamental advantages of considering response distributions over
summary statistic-based evaluations. Hoßfeld \textit{et al.} take this idea
further and show how to approximate response distributions given a QoS-to-MOS
mapping function~\cite{Hossfeld2020}. With our work we follow the trend of
response distribution analysis. At the same time, we indicate that interpretable GSD
model parameters can serve as summary statistics well describing underlying
response distribution.

Modelling individual responses generation process is another important thread of
MQA research focusing on response distribution analysis. The idea was first
proposed by Janowski and Pinson and termed
\textit{subject model}\footnote{The word \textit{subject}
refers to subjective experiment participant.}~\cite{Janowski2015}.
Li and Bampis took on the approach and proposed an extended subject
model~\cite{Li2017}. In their formulation of the model they considered subject bias,
subject inconsistency and stimulus ambiguity.
Reference \cite{Li2020EI} proposes an updated,
simpler version of the same model. Authors of~\cite{Li2020EI} convincingly show
that their model addresses shortcomings of subjective data analysis methods
put forward in several MQA-related ITU recommendations. Our work extends this
arc of research. We model individual responses generation process with the
Generalised Score Distribution (GSD) model. The model first came to light
in~\cite{Nawala2020ACM}, where we showed how it could be applied to check
subjective responses consistency.

We are not the first ones to notice that subjective responses modelling approach should
reflect data it operates on. Specifically, both \cite{Li2020ACM} and \cite{Pezzulli2021}
propose models taking into account ordinal nature of subjective responses coming
from MQA experiments.

Considering the literature review presented,
our work is novel in two respects. First, our proposed GSD model is the first
two-parameter model to model the complete variance range
observable for subjective responses expressed on an ordinal scale.
Second, to the best of our knowledge, we are the first ones to show
that our subjective responses generation process modelling approach outperforms
the standard approach based on empirical distribution when it comes to bootstrapping.


\section{Methodology}
\label{sec:methodology}
In this section we describe the methodology we use to substantiate the claims made
in the introduction. Section~\ref{ssec:subjective_response_as_a_random_variable}
describes our idea of treating subjective responses from MQA experiments as realisations
of a discrete random variable. Section~\ref{ssec:g_test_and_p_p_plot} shows how
we test the goodness-of-fit of the models we take into account. It also
presents how we interpret
resulting $p$-values. Section~\ref{ssec:data_sets} highlights data sets we use
to test the GSD on real data. Finally, Sec.~\ref{ssec:bootstrapping} details the
procedure we use to test GSD's performance when it comes to subjective responses
bootstrapping.

\subsection{Subjective Response as a Random Variable}
\label{ssec:subjective_response_as_a_random_variable}
We propose to think about responses from MQA subjective experiments as realisations
of a discrete random variable $U$. Since we focus on responses expressed on the 5-level
Absolute Category Rating (ACR) scale (cf. Sec. 6.1 of~\cite{itutp910}), $U$ can take
values from the $\{1, 2, 3, 4, 5\}$ set. To make the distribution of $U$ practically
useful, we need to parametrise it. Our experiences show that distributions with
one parameter do not properly fit real data. Thus, we focus on two-parameter distributions.
The following shows a general formulation of such distributions
\begin{equation}
    \label{eq:u_cdf}
    U \sim F(\lambda, \theta),
\end{equation}
where $F()$ is a cumulative distribution function, $\lambda$ is a parameter describing
central tendency of the distribution and $\theta$ expresses distribution spread.
Importantly, we assume that $F()$ reflects response distribution of each stimulus
in a subjective experiment. Per-stimulus values of $\lambda$ and $\theta$ define the
exact shape of $F()$.

Now, there are at least two approaches to proposing the exact formulation of $F()$.
The first one (which is more popular in the MQA literature) is to assume that subjective
responses follow a continuous normal distribution. The second one (which we take in
this paper) is to assume responses follow a discrete distribution and, more precisely,
the Generalised Score Distribution (GSD).\footnote{Although GSD's name refers to
scores, we use the word ``responses'' to refer to opinions formulated by observers taking
part in a MQA experiment.}

The approach assuming subjective responses follow continuous normal distribution is
best described by introducing an intermediate continuous random variable
$O \sim \mathcal{N}(\mu, \sigma^2)$, where $\mu$ describes the mean and $\sigma^2$ the variance
of the normal distribution. Since $U$ is discrete and $O$ is continuous, we need to
introduce a mapping between the two. In other words, $O$ must be discretized and
censored as follows
\begin{equation}\label{eq:discretise_234}
P(U = s) = \int_{s-0.5}^{s+0.5}\frac{1}{2\pi \sigma}e^{-\frac{(o-\mu)^2}{2\sigma^2}} do
\end{equation}
for $s = \{2,3,4\}$ and 
\begin{equation}\label{eq:discretise_1}
P(U = 1) = \int_{-\infty}^{1.5}\frac{1}{2\pi \sigma}e^{-\frac{(o-\mu)^2}{2\sigma^2}} do,
\end{equation}
\begin{equation}\label{eq:discretise_5}
P(U = 5) = \int_{4.5}^{\infty}\frac{1}{2\pi \sigma}e^{-\frac{(o-\mu)^2}{2\sigma^2}} do.
\end{equation}
Such a construct (i.e., a thresholded cumulative normal distribution) is quite popular
in latent variable analysis~\cite{Becker1992}. Thus, we follow the appropriate nomenclature
and refer to this model as Ordered Probit.
\jntodo{Mention the paper from 70' recommended by Pablo. This paper was the first
one to mention the Ordered Probit model.}

\subsubsection{Generalised Score Distribution}
Our approach to modelling subjective responses does not require any mapping between
an intermediate random variable and $U$. This is because the GSD already is a discrete
distribution. Thus, we can directly write $U \sim GSD(\psi, \rho)$, where $\psi$
expresses so called true quality and $\rho$ expresses responses spread. The true
quality parameter $\psi$ can be intuitively understood as a mean response for a given
stimulus, if we were to ask for opinion the complete population of observers. Contrary
to Ordered Probit's $\mu$, $\psi$ reflects the 5-level ACR scale and is bounded to the
$[1, 5]$ range.\footnote{To make the discussion easy to comprehend, we limit ourselves
to the version of the GSD reflecting a 5-level scale. However, the GSD can
describe any discrete process with domain of size $M$, where $M$ is a natural number
greater than 1.}
The other GSD's parameter, $\rho$, is a linear function of $V(U)$
(i.e., variance of $U$). Furthermore, $\rho$ is bounded to the $[0, 1]$ interval and
expresses what portion of possible variance is present in realisations of $U$.
Please note here that any discrete distribution with a limited domain
(e.g., $U \sim F(\lambda, \theta)$)
has its mean
value $E(U)$ and variance $V(U)$ bounded (cf. Fig.~\ref{fig:ghost_figures}). One more important property of $\rho$ is that it represents
responses confidence. In other words, it is inversely proportional to variance observed
in responses (the higher the observed variance, the lower the value of $\rho$). Yet
another way to put it is to say that the greater the value of $\rho$, the closer to
$\psi$ observed responses are.
\jncomment{If the GSD can describe any $M$-point scale, what is the complete set of $M$s?
Is this the set of all natural numbers?}
Importantly, the GSD is able to model the complete range
of possible variances for a given $M$-point scale (with $M \in \mathbb{N}: M > 1$).
For more details regarding the GSD we refer the reader to~\cite{Janowski2019arxiv}.

To make GSD's description more concrete, let us take a look at its internal structure.
We start by showing a more detailed form of the $U \sim GSD(\psi, \rho)$ expression:
\begin{equation}
U \sim \psi + \epsilon, \label{eq:gsd}
\end{equation}
where $\epsilon$ expresses uncertainty regarding mean response represented by $\psi$.
$\psi$ is one constant number estimated for a stimulus of interest.
Notice that $\psi = E(U)$.
$\epsilon$, on the
other hand, follows a distribution parameterised with a single parameter $\rho$. What is
more, $\epsilon$'s distribution satisfies the following two criteria: (i) its mean equals
zero and (ii) its variance is a linear function of $\rho$. In Appendix~\ref{app:full_gsd}
(see the supplemental material)
we show the exact formulation of $\epsilon$'s distribution. Here, we only mention that
this distribution is a mixture of the following distributions: binomial, beta-binomial
and one- or two-point distribution (whether one- or two-point distribution is used depends
on the value of $\psi$). Importantly, we reparameterise the
distributions in the mixture to make them satisfy the two criteria that $\epsilon$'s
distribution must follow. As a result, the reparameterised distributions in the mixture
depend only on a single parameter $\rho$.

\begin{figure*}[!t]
    \centering
    \subfloat[]{\resizebox{.3\textwidth}{!}{
\begin{tikzpicture}

\definecolor{color0}{rgb}{0.886274509803922,0.290196078431373,0.2}
\definecolor{color1}{rgb}{0.203921568627451,0.541176470588235,0.741176470588235}
\definecolor{color2}{rgb}{0.596078431372549,0.556862745098039,0.835294117647059}
\definecolor{color3}{rgb}{0.984313725490196,0.756862745098039,0.368627450980392}
\definecolor{color4}{rgb}{0.556862745098039,0.729411764705882,0.258823529411765}
\definecolor{color5}{rgb}{0.75,0,0.75}

\begin{axis}[
axis background/.style={fill=white},
axis line style={white!89.80392156862746!black},
legend cell align={left},
legend entries={\hspace{-.2cm}\textbf{$\rho$},{0.95},{0.88},{0.81},{0.72},{0.61},{0.38}},
legend style={draw=white!80.0!black, fill=white},
tick align=outside,
tick pos=left,
x grid style={white!89.80392156862746!black},
xlabel={Score $s$},
xmajorgrids,
xmin=0.8, xmax=5.2,
y grid style={white!89.80392156862746!black},
ylabel={$P(U = s)$},
ymajorgrids,
ymin=-0.05, ymax=1.05
]
\addlegendimage{empty legend}
\addplot [line width=0.08000000000000002pt, color0, dashed, mark=*, mark size=4, mark options={solid}]
table [row sep=\\]{%
1	0.72139609375 \\
2	0.258290625 \\
3	0.0192515625 \\
4	0.001040625 \\
5	2.109375e-05 \\
};
\addplot [line width=0.08000000000000002pt, color1, dashed, mark=triangle*, mark size=4, mark options={solid,rotate=180}]
table [row sep=\\]{%
1	0.748496941750038 \\
2	0.208222542278198 \\
3	0.038350722665177 \\
4	0.00464316083489919 \\
5	0.000286632471687573 \\
};
\addplot [line width=0.08000000000000002pt, color2, dashed, mark=diamond*, mark size=4, mark options={solid}]
table [row sep=\\]{%
1	0.77096417921922 \\
2	0.171207946584666 \\
3	0.0460910849306944 \\
4	0.0103372735077423 \\
5	0.00139951575767982 \\
};
\addplot [line width=0.08000000000000002pt, white!46.666666666666664!black, dashed, mark=x, mark size=4, mark options={solid}]
table [row sep=\\]{%
1	0.795829157800537 \\
2	0.134191333399622 \\
3	0.0484510529979119 \\
4	0.017207262603162 \\
5	0.00432119319876703 \\
};
\addplot [line width=0.08000000000000002pt, color3, dashed, mark=pentagon*, mark size=4, mark options={solid}]
table [row sep=\\]{%
1	0.821895746624118 \\
2	0.0996567312454621 \\
3	0.0451553265189036 \\
4	0.0231361667293331 \\
5	0.0101560288821829 \\
};
\addplot [line width=0.08000000000000002pt, color4, dashed, mark=square*, mark size=4, mark options={solid}]
table [row sep=\\]{%
1	0.866684689952905 \\
2	0.0497367503924647 \\
3	0.0296816091051805 \\
4	0.0246877708006279 \\
5	0.0292091797488226 \\
};
\addplot [very thick, color5, forget plot]
table [row sep=\\]{%
1.3	0 \\
1.3	1 \\
};
\node at (axis cs:1.4,0.8)[
  anchor=base west,
  text=black,
  rotate=0.0
]{ $\psi = 1.30$};
\end{axis}

\end{tikzpicture}}%
        \label{fig:gsd_example_a}}
    \hfil
    \subfloat[]{\resizebox{.3\textwidth}{!}{
\begin{tikzpicture}

\definecolor{color0}{rgb}{0.886274509803922,0.290196078431373,0.2}
\definecolor{color1}{rgb}{0.203921568627451,0.541176470588235,0.741176470588235}
\definecolor{color2}{rgb}{0.596078431372549,0.556862745098039,0.835294117647059}
\definecolor{color3}{rgb}{0.984313725490196,0.756862745098039,0.368627450980392}
\definecolor{color4}{rgb}{0.556862745098039,0.729411764705882,0.258823529411765}
\definecolor{color5}{rgb}{0.75,0,0.75}

\begin{axis}[
axis background/.style={fill=white},
axis line style={white!89.80392156862746!black},
legend cell align={left},
legend entries={\hspace{-.2cm}\textbf{$\rho$},{0.95},{0.88},{0.81},{0.72},{0.61},{0.38}},
legend style={draw=white!80.0!black, fill=white},
tick align=outside,
tick pos=left,
x grid style={white!89.80392156862746!black},
xlabel={Score $s$},
xmajorgrids,
xmin=0.8, xmax=5.2,
y grid style={white!89.80392156862746!black},
ylabel={$P(U = s)$},
ymajorgrids,
ymin=-0.05, ymax=1.05
]
\addlegendimage{empty legend}
\addplot [line width=0.08000000000000002pt, color0, dashed, mark=*, mark size=4, mark options={solid}]
table [row sep=\\]{%
1	0.0605281332818022 \\
2	0.794662643551237 \\
3	0.130343269655477 \\
4	0.0132129969081272 \\
5	0.00125295660335689 \\
};
\addplot [line width=0.08000000000000002pt, color1, dashed, mark=triangle*, mark size=4, mark options={solid,rotate=180}]
table [row sep=\\]{%
1	0.145267519876325 \\
2	0.647190344522968 \\
3	0.172823847173145 \\
4	0.0317111925795053 \\
5	0.00300709584805654 \\
};
\addplot [line width=0.08000000000000002pt, color2, dashed, mark=diamond*, mark size=4, mark options={solid}]
table [row sep=\\]{%
1	0.230006906470848 \\
2	0.4997180454947 \\
3	0.215304424690813 \\
4	0.0502093882508834 \\
5	0.00476123509275618 \\
};
\addplot [line width=0.08000000000000002pt, white!46.666666666666664!black, dashed, mark=x, mark size=4, mark options={solid}]
table [row sep=\\]{%
1	0.316802044146678 \\
2	0.37043976700906 \\
3	0.221868434092756 \\
4	0.0777356542005991 \\
5	0.0131541005509078 \\
};
\addplot [line width=0.08000000000000002pt, color3, dashed, mark=pentagon*, mark size=4, mark options={solid}]
table [row sep=\\]{%
1	0.394171811991077 \\
2	0.285141337039146 \\
3	0.184045116936097 \\
4	0.0997985070460573 \\
5	0.0368432269876218 \\
};
\addplot [line width=0.08000000000000002pt, color4, dashed, mark=square*, mark size=4, mark options={solid}]
table [row sep=\\]{%
1	0.532202936822452 \\
2	0.153334274334812 \\
3	0.107779556060854 \\
4	0.0956263175840495 \\
5	0.111056915197833 \\
};
\addplot [very thick, color5, forget plot]
table [row sep=\\]{%
2.1	0 \\
2.1	1 \\
};
\node at (axis cs:2.2,0.8)[
  anchor=base west,
  text=black,
  rotate=0.0
]{ $\psi = 2.10$};
\end{axis}

\end{tikzpicture}}%
        \label{fig:gsd_example_b}}
    \hfil
    \subfloat[]{\resizebox{.3\textwidth}{!}{
\begin{tikzpicture}

\definecolor{color0}{rgb}{0.886274509803922,0.290196078431373,0.2}
\definecolor{color1}{rgb}{0.203921568627451,0.541176470588235,0.741176470588235}
\definecolor{color2}{rgb}{0.596078431372549,0.556862745098039,0.835294117647059}
\definecolor{color3}{rgb}{0.984313725490196,0.756862745098039,0.368627450980392}
\definecolor{color4}{rgb}{0.556862745098039,0.729411764705882,0.258823529411765}
\definecolor{color5}{rgb}{0.75,0,0.75}

\begin{axis}[
axis background/.style={fill=white},
axis line style={white!89.80392156862746!black},
legend cell align={left},
legend entries={\hspace{-.2cm}\textbf{$\rho$},{0.95},{0.88},{0.81},{0.72},{0.61},{0.38}},
legend style={draw=white!80.0!black, fill=white},
tick align=outside,
tick pos=left,
x grid style={white!89.80392156862746!black},
xlabel={Score $s$},
xmajorgrids,
xmin=0.8, xmax=5.2,
y grid style={white!89.80392156862746!black},
ylabel={$P(U = s)$},
ymajorgrids,
ymin=-0.05, ymax=1.05
]
\addlegendimage{empty legend}
\addplot [line width=0.08000000000000002pt, color0, dashed, mark=*, mark size=4, mark options={solid}]
table [row sep=\\]{%
1	0.0185348096301595 \\
2	0.18048492784562 \\
3	0.743586358682183 \\
4	0.0472332605781363 \\
5	0.0101606432639014 \\
};
\addplot [line width=0.08000000000000002pt, color1, dashed, mark=triangle*, mark size=4, mark options={solid,rotate=180}]
table [row sep=\\]{%
1	0.0444835431123828 \\
2	0.223163826829488 \\
3	0.594607260837239 \\
4	0.113359825387527 \\
5	0.0243855438333634 \\
};
\addplot [line width=0.08000000000000002pt, color2, dashed, mark=diamond*, mark size=4, mark options={solid}]
table [row sep=\\]{%
1	0.0704322765946062 \\
2	0.265842725813356 \\
3	0.445628162992295 \\
4	0.179486390196918 \\
5	0.0386104444028254 \\
};
\addplot [line width=0.08000000000000002pt, white!46.666666666666664!black, dashed, mark=x, mark size=4, mark options={solid}]
table [row sep=\\]{%
1	0.114405372370277 \\
2	0.276308916717531 \\
3	0.323641015625745 \\
4	0.216169729114808 \\
5	0.0694749661716385 \\
};
\addplot [line width=0.08000000000000002pt, color3, dashed, mark=pentagon*, mark size=4, mark options={solid}]
table [row sep=\\]{%
1	0.178081405530565 \\
2	0.244367463186289 \\
3	0.249159177257745 \\
4	0.206253633803385 \\
5	0.122138320222017 \\
};
\addplot [line width=0.08000000000000002pt, color4, dashed, mark=square*, mark size=4, mark options={solid}]
table [row sep=\\]{%
1	0.313469658678537 \\
2	0.158680186692613 \\
3	0.136641487850938 \\
4	0.146797829506136 \\
5	0.244410837271775 \\
};
\addplot [very thick, color5, forget plot]
table [row sep=\\]{%
2.85	0 \\
2.85	1 \\
};
\node at (axis cs:2.95,0.8)[
  anchor=base west,
  text=black,
  rotate=0.0
]{ $\psi = 2.85$};
\end{axis}

\end{tikzpicture}}%
        \label{fig:gsd_example_c}}
    \caption{Realisations (in the form of probability mass functions) of the GSD for
    a 5-point scale and various values of
    parameters $\psi$ and $\rho$. Notice how the growing value of $\rho$ corresponds to
    more responses accumulating close to the value of $\psi$.}
    \label{fig:gsd_example}
\end{figure*}
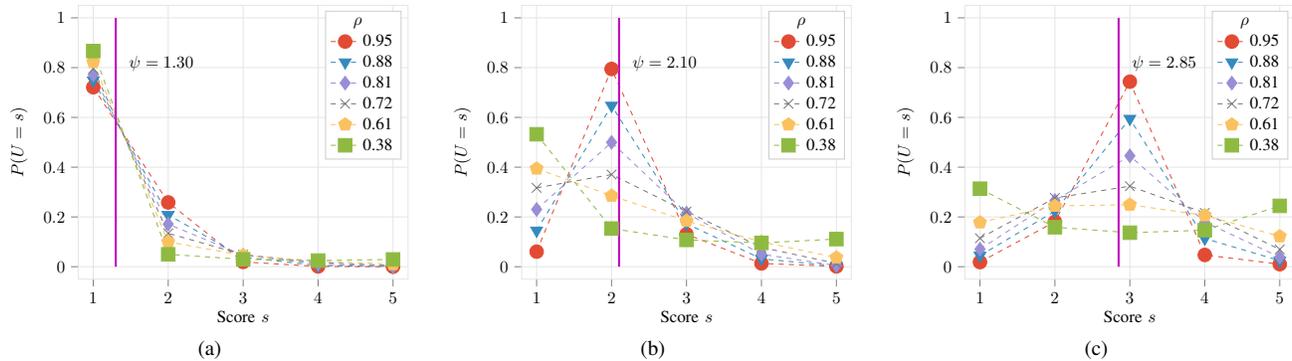

Fig.~\ref{fig:gsd_example} shows various
realisations of the GSD for different values of $\psi$ and $\rho$.
Please notice how flexible the GSD is. For example, in Fig.~\ref{fig:gsd_example_c},
for the case of $\rho = 0.38$, the GSD takes the form of a distribution with two modes
(one mode at response 1 and another at response 5). Apart from this extreme example,
GSD's shape follows common sense intuition regarding response distributions observed
in typical MQA subjective experiments.

\subsection{G-test and P--P Plot}
\label{ssec:g_test_and_p_p_plot}
In order to validate if a distribution (or a model) fits specific data we have to perform a
two-step
procedure. The first step is to estimate distribution parameters for a sample of interest.
The second step is to test a null hypothesis stating that the sample truly comes from
the assumed distribution (GSD or Ordered Probit in our case),
given the parameters estimated in the first step.
We use a standard likelihood ratio approach to test the goodness-of-fit (GoF) of the
two models (the GSD and Ordered Probit).
More precisely, we use the so called G-test of GoF (cf. Sec. 14.3.4 of~\cite{Agresti}). 
Because sample sizes we consider are predominantly small, we do not use the asymptotic
distribution for calculating the $p$-value. On the contrary, we estimate the $p$-value
utilising a bootstrapped version of the G-test (see Appendix \ref{app:gtest} in the
supplemental material).
(For broader theoretical considerations on the topic please take a look at~\cite{Efron}.)

Since each MQA subjective experiment we analyse contains multiple stimuli, we need to
perform the G-test multiple times (as many as there are stimuli in the experiment).
The result of each G-test is a $p$-value. This means we get a vector of $p$-values for
each experiment we take into account. To be able to efficiently draw conclusions
regarding a vector of $p$-values we use $p$-value P--P plots (where P--P stands for
probability--probability)~\cite{Schweder1982}. For a detailed discussion regarding
$p$-value P--P plots for the GSD we refer the reader to~\cite{Nawala2020ACM}.

\subsection{Data Sets}
\label{ssec:data_sets}
To test the GSD in practice we make use of more than one million individual subjective
responses (exactly $1\,183\,696$). We take into account responses coming from 33
subjective experiments that assess quality (or other traits) of more than nine thousand
stimuli (exactly
$9\,290$). Table~\ref{tab:data_sets_summary} presents an overview of data sets we use.
Importantly, we classify data sets into three types: (i) typical, (ii) broadly
understood and (iii) non-MQA. The types reflect how much a given data set follows
best practices and recommendations regarding organising MQA experiments. \textit{Typical}
experiments tightly follow best practices and recommendations. \textit{Broadly understood}
experiments follow these best practices and recommendations generally, but deviate from
them in some aspects. Finally, \textit{non-MQA} experiments are not MQA experiments at all.
We include these to check GSD's performance on data outside of GSD's intended application
scope. Please note that one data set may correspond to multiple experiments (cf. the
``No. of Exp.'' row of Table~\ref{tab:data_sets_summary}). For example,
the MM2 data set consists of 10 separate experiments. Thus, although we use data from
11 data sets, they amount to 33 experiments.
\begin{table*}[!t]
    \centering
    \begin{threeparttable}
        \renewcommand{\arraystretch}{1.3}
        \caption{An Overview of Data Sets We Use to Test the GSD on Real Data}
        \label{tab:data_sets_summary}
        \begin{tabular}{p{1.6cm}p{.7cm}p{.8cm}p{.8cm}p{.8cm}p{.9cm}p{.9cm}p{1cm}p{1.1cm}p{1.1cm}p{1.2cm}p{1.2cm}}
        \hline\hline
        Study &
            ITU \cite{itutsupp23}  &
            HDTV \cite{HDTV_Phase_I_test}  &
            MM2 \cite{Pinson2012}  &
            14-505 \cite{AGH_NTIA_14-505} &
            ITS4S \cite{ITS4S} &
            NFLX &
            ITERO \cite{Perez2021} &
            ITS4S2 \cite{ITS4S2} &
            Naderi \cite{Naderi2020} &
            MovieLens \cite{Harper2015} &
            Personality \cite{Nguyen2018} \\ \hline
        Year &
            1995 &
            2010  &
            2012 &
            2014   &
            2018  &
            2018 &
            2019 &
            2019   &
            2020  &
            2003      &
            2018        \\
        No. of Exp.   &
            1    &
            6     &
            10   &
            1      &
            2     &
            4    &
            3    &
            1      &
            3     &
            1         &
            1           \\
        Total No. of Stimuli &
            176  &
            1 008 &
            600  &
            114    &
            1 025 &
            720  &
            330  &
            1 429  &
            170   &
            3 708     &
            10          \\
        Total No. of Responses &
          4 224 &
          24 192 &
          12 780 &
          7 076 &
          26 926 &
          36 000 &
          25 080 &
          22 864 &
          22 511 &
          $1\,000\,209$ &
          1 834 \\
        Type &
          typical &
          typical &
          typical &
          typical &
          typical &
          typical &
          bu &
          bu &
          bu &
          non-MQA &
          non-MQA \\
        Stimulus type &
          speech &
          video &
          av &
          video &
          video &
          video &
          video &
          image &
          speech &
          mr &
          mr \\ \hline\hline
        \end{tabular}
        \begin{tablenotes}
            \item Exp. stands for experiments; av stands for audiovisual; mr stands for
            movie recommendation; bu stands for broadly understood.
        \end{tablenotes}
    \end{threeparttable}
\end{table*}

We do not provide detailed descriptions of the data sets here. Instead, in
Table~\ref{tab:data_sets_summary} we link to references describing each data set.
The only exception to this rule is the NFLX data set. Since its description has not yet
been published, we describe the data set briefly.

Experiments included in the NFLX data set investigated influence of per-scene quality
changes on opinion of
human observers. 200 observers assessed quality of 320
stimuli.\footnotemark~Ten seconds long video clips
(without audio) were used as stimuli. The clips had a resolution of 1920x1080 pixels.
Quality degradations were applied solely through
video compression. However, since per-scene compression was used, quality switches
occurred during playback as well.
Contents spanned a wide range of categories and were taken from
Netflix's catalogue. This made the experiments more ecologically valid, but also meant
the clips could not be publicly shared. The clips were displayed on either a TV
or a tablet (both with the native resolution of 1920x1080 pixels).
What is more, some participants were asked to provide their opinion during
video playback. They were encouraged to use a software slider displayed at the bottom of the screen.
In total, four experiments were performed: (i) with the TV and the software slider,
(ii) with the TV and without the slider, (iii) with the tablet and the slider and
(iv) with the tablet and without the slider. Participants were recruited through a
temporary working agency. Care was taken not to over-represent the 18 to 25 age group.
All four experiments took place in a controlled environment and were generally following
provisions of Rec. ITU-T P.913~\cite{P913_2021}. The experiments were
performed in accordance with the Absolute Category Rating with Hidden Reference (ACR-HR)
method (cf. Sec. 7.2.2 of \cite{P913_2021}). Thus, participants provided
their responses using the 5-level ACR scale.
\footnotetext{In Table~\ref{tab:data_sets_summary} we write about 720 stimuli in this
data set, since we treat each of the four experiments as separate. Because each
of the four experiments investigated 180 stimuli, we end up with 720 stimuli
in total.}

\subsection{Bootstrapping}
\label{ssec:bootstrapping}
\jntodo{Consider rewriting it to make it easily digestible.
In other words, leave the nasty parts for the supplement or appendix.}
To compare GSD's generalisability to that of the empirical
distribution (which is typically used for resampling),
we introduce the following procedure. We start by
generating $MC$ (e.g., $MC = 10\,000$) bootstrap samples
from the empirical probability mass function (EPMF)
of the large sample. Importantly, we generate bootstrap
samples with significantly fewer
observations than those in the large sample (e.g., $n=24$ observations
in each bootstrap sample for $N=200$ observations in the large sample).
Now, we fit the GSD to each $r$-th bootstrap
sample. This yields estimates of each response category probability
$\left( \hat{q}^r_1, \hat{q}^r_2, \hat{q}^r_3, \hat{q}^r_4, \hat{q}^r_5 \right)$.
We use those estimates to calculate the likelihood function for the
large sample $\mathcal{L}^r_{GSD}$.
We repeat the procedure for each bootstrap sample, but this time
using the EPMF of the bootstrap sample
to find response category probability
estimates $\left( \hat{v}^r_1, \hat{v}^r_2, \hat{v}^r_3, \hat{v}^r_4,
\hat{v}^r_5 \right)$.
Having the likelihoods for both the GSD ($\mathcal{L}^r_{GSD}$)
and empirical distribution ($\mathcal{L}^r_{e}$)
we introduce a statistic $W_r$ based on the quotient of the two values.
In other words, we introduce a statistic based on the likelihood
ratio: $W_r = \ln \left( \mathcal{L}^r_{GSD} / \mathcal{L}^r_e \right)$.
Value of the quotient signifies which approach better describes the
large sample. (Note that there are as many quotients as there are
bootstrap samples.) Now, we use the quotients to estimate the probability
$p_{\mathrm{GSD}}$
that the GSD model-based estimates of response category probabilities 
in the large sample yield higher likelihood function value
($\mathcal{L}_{GSD}$) than
the likelihood function value we get if we use the EPMF-based
estimates ($\mathcal{L}_{e}$).
We also do the same for the empirical
distribution and estimate the probability
that the EPMF-based estimates yield higher
likelihood function value than that yielded by the GSD model-based
estimates and denote this probability by $p_{\mathrm{e}}$.
We calculate the 95\% confidence interval for
$p_{\mathrm{GSD}}-p_{\mathrm{e}}=P(W_r > 0)-P(W_r < 0)$ and
denote its lower (or left) bound as $L$ and upper (or right) bound as $R$. If
$L > 0$ then the GSD performs better than the empirical distribution.
If $R < 0$ then the empirical
distribution performs better. If $[L, R]$ contains zero, there
is no significant difference between the GSD and empirical distribution.
We provide the precise description of the procedure given above
in Appendix~\ref{app:generalisability_test} (see the supplemental material).

Since we use the subsample to make inferences about the large sample,
there is a risk of overfitting. Put differently, by fitting any model
too precisely to the subsample we are at risk of finding model parameter
estimates that are suboptimal from the point of view of the large sample.
This is because the subsample represents only limited information about the
large sample. Intuitively, we should not entirely trust the data we
observe in the subsample.
To address the issue we apply parameter estimation modification that
prevents probability estimators we use from yielding response category
probabilities equal to $0$ (for any response category).
In other words, we expect that, at the population
level, there is no response category that would be assigned no
observations (even if the estimation result for the subsample says
something else). This results in modified estimation procedures for both
the GSD and empirical distribution.
The detailed estimation correction procedures we use are described in
Appendix~\ref{app:parameters_estimation_modification} (see the supplemental
material).


\section{Results}
\label{sec:results}
We present here results reflecting our contributions mentioned in the introduction.
Sec.~\ref{ssec:interpretable_parameters} puts forward evidence supporting the claim
that the GSD has easy to interpret parameters.
Sec.~\ref{ssec:good_description_of_typical_mqa_experiments} shows that the GSD
well describes response distributions from typical MQA experiments. It also indicates
that GSD does not perform well for atypical MQA and non-MQA experiments.
Sec.~\ref{ssec:better_than_empirical_distribution} reveals that the GSD outperforms
empirical distribution when it comes to subjective responses bootstrapping.
At last, Sec.~\ref{ssec:comparison_w_sota_model} evidences that
the GSD outperforms the state-of-the-art model both in terms of
goodness-of-fit testing and bootstrapping.

\subsection{Interpretable Parameters}
\label{ssec:interpretable_parameters}
\begin{figure*}[!t]
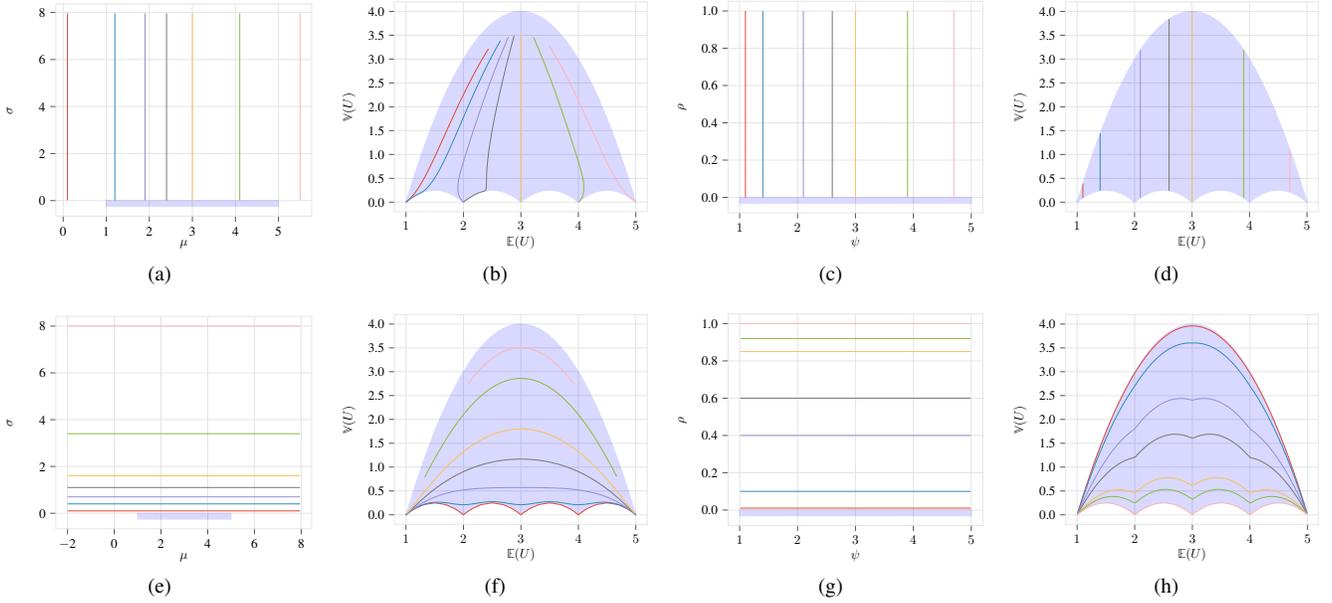

    \centering
    \subfloat[]{\resizebox{0.23\textwidth}{!}{\input{plots/space_mu.tex}}%
        \label{fig:ghost_ordered_probit_a}}
    \hfil
    \subfloat[]{\resizebox{0.23\textwidth}{!}{\input{plots/ghost_mu.tex}}%
        \label{fig:ghost_ordered_probit_b}}
    \hfil
    \subfloat[]{\resizebox{0.23\textwidth}{!}{
\begin{tikzpicture}

\definecolor{color0}{rgb}{0.886274509803922,0.290196078431373,0.2}
\definecolor{color1}{rgb}{0.203921568627451,0.541176470588235,0.741176470588235}
\definecolor{color2}{rgb}{0.596078431372549,0.556862745098039,0.835294117647059}
\definecolor{color3}{rgb}{0.984313725490196,0.756862745098039,0.368627450980392}
\definecolor{color4}{rgb}{0.556862745098039,0.729411764705882,0.258823529411765}
\definecolor{color5}{rgb}{1,0.709803921568627,0.72156862745098}

\begin{axis}[
axis background/.style={fill=white},
axis line style={white!89.8039215686275!black},
tick align=outside,
tick pos=left,
x grid style={white!89.8039215686275!black},
xlabel={\(\displaystyle \psi\)},
xmajorgrids,
xmin=0.8, xmax=5.2,
xtick style={color=white!33.3333333333333!black},
y grid style={white!89.8039215686275!black},
ylabel={\(\displaystyle \rho\)},
ymajorgrids,
ymin=-0.0828125, ymax=1.0515625,
ytick style={color=white!33.3333333333333!black},
ytick={-0.2,0,0.2,0.4,0.6,0.8,1,1.2},
yticklabels={\ensuremath{-}0.2,0.0,0.2,0.4,0.6,0.8,1.0,1.2}
]
\path [draw=blue, fill=blue, opacity=0.15, very thin]
(axis cs:1,0)
--(axis cs:1,-0.03125)
--(axis cs:5,-0.03125)
--(axis cs:5,0)
--(axis cs:5,0)
--(axis cs:1,0)
--cycle;

\addplot [semithick, color0]
table {%
1.1 0
1.1 0.05
1.1 0.1
1.1 0.15
1.1 0.2
1.1 0.25
1.1 0.3
1.1 0.35
1.1 0.4
1.1 0.45
1.1 0.5
1.1 0.55
1.1 0.6
1.1 0.65
1.1 0.7
1.1 0.75
1.1 0.8
1.1 0.85
1.1 0.9
1.1 0.95
1.1 1
};
\addplot [semithick, color1]
table {%
1.4 0
1.4 0.05
1.4 0.1
1.4 0.15
1.4 0.2
1.4 0.25
1.4 0.3
1.4 0.35
1.4 0.4
1.4 0.45
1.4 0.5
1.4 0.55
1.4 0.6
1.4 0.65
1.4 0.7
1.4 0.75
1.4 0.8
1.4 0.85
1.4 0.9
1.4 0.95
1.4 1
};
\addplot [semithick, color2]
table {%
2.1 0
2.1 0.05
2.1 0.1
2.1 0.15
2.1 0.2
2.1 0.25
2.1 0.3
2.1 0.35
2.1 0.4
2.1 0.45
2.1 0.5
2.1 0.55
2.1 0.6
2.1 0.65
2.1 0.7
2.1 0.75
2.1 0.8
2.1 0.85
2.1 0.9
2.1 0.95
2.1 1
};
\addplot [semithick, white!46.6666666666667!black]
table {%
2.6 0
2.6 0.05
2.6 0.1
2.6 0.15
2.6 0.2
2.6 0.25
2.6 0.3
2.6 0.35
2.6 0.4
2.6 0.45
2.6 0.5
2.6 0.55
2.6 0.6
2.6 0.65
2.6 0.7
2.6 0.75
2.6 0.8
2.6 0.85
2.6 0.9
2.6 0.95
2.6 1
};
\addplot [semithick, color3]
table {%
3 0
3 0.05
3 0.1
3 0.15
3 0.2
3 0.25
3 0.3
3 0.35
3 0.4
3 0.45
3 0.5
3 0.55
3 0.6
3 0.65
3 0.7
3 0.75
3 0.8
3 0.85
3 0.9
3 0.95
3 1
};
\addplot [semithick, color4]
table {%
3.9 0
3.9 0.05
3.9 0.1
3.9 0.15
3.9 0.2
3.9 0.25
3.9 0.3
3.9 0.35
3.9 0.4
3.9 0.45
3.9 0.5
3.9 0.55
3.9 0.6
3.9 0.65
3.9 0.7
3.9 0.75
3.9 0.8
3.9 0.85
3.9 0.9
3.9 0.95
3.9 1
};
\addplot [semithick, color5]
table {%
4.7 0
4.7 0.05
4.7 0.1
4.7 0.15
4.7 0.2
4.7 0.25
4.7 0.3
4.7 0.35
4.7 0.4
4.7 0.45
4.7 0.5
4.7 0.55
4.7 0.6
4.7 0.65
4.7 0.7
4.7 0.75
4.7 0.8
4.7 0.85
4.7 0.9
4.7 0.95
4.7 1
};
\end{axis}

\end{tikzpicture}}%
        \label{fig:ghost_gsd_a}}
    \hfil
    \subfloat[]{\resizebox{0.23\textwidth}{!}{
\begin{tikzpicture}

\definecolor{color0}{rgb}{0.886274509803922,0.290196078431373,0.2}
\definecolor{color1}{rgb}{0.203921568627451,0.541176470588235,0.741176470588235}
\definecolor{color2}{rgb}{0.596078431372549,0.556862745098039,0.835294117647059}
\definecolor{color3}{rgb}{0.984313725490196,0.756862745098039,0.368627450980392}
\definecolor{color4}{rgb}{0.556862745098039,0.729411764705882,0.258823529411765}
\definecolor{color5}{rgb}{1,0.709803921568627,0.72156862745098}

\begin{axis}[
axis background/.style={fill=white},
axis line style={white!89.8039215686275!black},
tick align=outside,
tick pos=left,
x grid style={white!89.8039215686275!black},
xlabel={\(\displaystyle \mathbb{E}(U)\)},
xmajorgrids,
xmin=0.8, xmax=5.2,
xtick style={color=white!33.3333333333333!black},
y grid style={white!89.8039215686275!black},
ylabel={\(\displaystyle \mathbb{V}(U)\)},
ymajorgrids,
ymin=-0.200000000000015, ymax=4.2,
ytick style={color=white!33.3333333333333!black},
ytick={-0.5,0,0.5,1,1.5,2,2.5,3,3.5,4,4.5},
yticklabels={\ensuremath{-}0.5,0.0,0.5,1.0,1.5,2.0,2.5,3.0,3.5,4.0,4.5}
]
\path [draw=blue, fill=blue, opacity=0.15, very thin]
(axis cs:1,0)
--(axis cs:1,0)
--(axis cs:1.05,0.0475)
--(axis cs:1.1,0.0900000000000001)
--(axis cs:1.15,0.1275)
--(axis cs:1.2,0.16)
--(axis cs:1.25,0.1875)
--(axis cs:1.3,0.21)
--(axis cs:1.35,0.2275)
--(axis cs:1.4,0.24)
--(axis cs:1.45,0.2475)
--(axis cs:1.5,0.25)
--(axis cs:1.55,0.2475)
--(axis cs:1.6,0.24)
--(axis cs:1.65,0.2275)
--(axis cs:1.7,0.21)
--(axis cs:1.75,0.1875)
--(axis cs:1.8,0.16)
--(axis cs:1.85,0.127499999999999)
--(axis cs:1.9,0.0899999999999994)
--(axis cs:1.95,0.0474999999999992)
--(axis cs:2,8.88178419700124e-16)
--(axis cs:2.05,0.0475000000000006)
--(axis cs:2.1,0.0900000000000008)
--(axis cs:2.15,0.127500000000001)
--(axis cs:2.2,0.160000000000001)
--(axis cs:2.25,0.1875)
--(axis cs:2.3,0.21)
--(axis cs:2.35,0.2275)
--(axis cs:2.4,0.24)
--(axis cs:2.45,0.2475)
--(axis cs:2.5,0.25)
--(axis cs:2.55,0.2475)
--(axis cs:2.6,0.24)
--(axis cs:2.65,0.2275)
--(axis cs:2.7,0.209999999999999)
--(axis cs:2.75,0.187499999999999)
--(axis cs:2.8,0.159999999999999)
--(axis cs:2.85,0.127499999999999)
--(axis cs:2.9,0.0899999999999987)
--(axis cs:2.95,0.0474999999999982)
--(axis cs:3,1.77635683940025e-15)
--(axis cs:3.05,0.0475000000000014)
--(axis cs:3.1,0.0900000000000015)
--(axis cs:3.15,0.127500000000002)
--(axis cs:3.2,0.160000000000001)
--(axis cs:3.25,0.187500000000001)
--(axis cs:3.3,0.210000000000001)
--(axis cs:3.35,0.227500000000001)
--(axis cs:3.4,0.24)
--(axis cs:3.45,0.2475)
--(axis cs:3.5,0.25)
--(axis cs:3.55,0.2475)
--(axis cs:3.6,0.24)
--(axis cs:3.65,0.227499999999999)
--(axis cs:3.7,0.209999999999999)
--(axis cs:3.75,0.187499999999999)
--(axis cs:3.8,0.159999999999999)
--(axis cs:3.85,0.127499999999998)
--(axis cs:3.9,0.0899999999999979)
--(axis cs:3.95,0.0474999999999974)
--(axis cs:4,2.66453525910037e-15)
--(axis cs:4.05,0.0475000000000022)
--(axis cs:4.1,0.0900000000000026)
--(axis cs:4.15,0.127500000000002)
--(axis cs:4.2,0.160000000000002)
--(axis cs:4.25,0.187500000000001)
--(axis cs:4.3,0.210000000000001)
--(axis cs:4.35,0.227500000000001)
--(axis cs:4.4,0.240000000000001)
--(axis cs:4.45,0.2475)
--(axis cs:4.5,0.25)
--(axis cs:4.55,0.2475)
--(axis cs:4.6,0.239999999999999)
--(axis cs:4.65,0.227499999999999)
--(axis cs:4.7,0.209999999999999)
--(axis cs:4.75,0.187499999999998)
--(axis cs:4.8,0.159999999999998)
--(axis cs:4.85,0.127499999999998)
--(axis cs:4.9,0.0899999999999969)
--(axis cs:4.95,0.0474999999999966)
--(axis cs:5,3.55271367880049e-15)
--(axis cs:5,-1.4210854715202e-14)
--(axis cs:5,-1.4210854715202e-14)
--(axis cs:4.95,0.197499999999985)
--(axis cs:4.9,0.389999999999985)
--(axis cs:4.85,0.577499999999988)
--(axis cs:4.8,0.759999999999988)
--(axis cs:4.75,0.937499999999988)
--(axis cs:4.7,1.10999999999999)
--(axis cs:4.65,1.27749999999999)
--(axis cs:4.6,1.43999999999999)
--(axis cs:4.55,1.59749999999999)
--(axis cs:4.5,1.74999999999999)
--(axis cs:4.45,1.89749999999999)
--(axis cs:4.4,2.03999999999999)
--(axis cs:4.35,2.17749999999999)
--(axis cs:4.3,2.30999999999999)
--(axis cs:4.25,2.43749999999999)
--(axis cs:4.2,2.55999999999999)
--(axis cs:4.15,2.67749999999999)
--(axis cs:4.1,2.78999999999999)
--(axis cs:4.05,2.89749999999999)
--(axis cs:4,2.99999999999999)
--(axis cs:3.95,3.09749999999999)
--(axis cs:3.9,3.19)
--(axis cs:3.85,3.2775)
--(axis cs:3.8,3.36)
--(axis cs:3.75,3.4375)
--(axis cs:3.7,3.51)
--(axis cs:3.65,3.5775)
--(axis cs:3.6,3.64)
--(axis cs:3.55,3.6975)
--(axis cs:3.5,3.75)
--(axis cs:3.45,3.7975)
--(axis cs:3.4,3.84)
--(axis cs:3.35,3.8775)
--(axis cs:3.3,3.91)
--(axis cs:3.25,3.9375)
--(axis cs:3.2,3.96)
--(axis cs:3.15,3.9775)
--(axis cs:3.1,3.99)
--(axis cs:3.05,3.9975)
--(axis cs:3,4)
--(axis cs:2.95,3.9975)
--(axis cs:2.9,3.99)
--(axis cs:2.85,3.9775)
--(axis cs:2.8,3.96)
--(axis cs:2.75,3.9375)
--(axis cs:2.7,3.91)
--(axis cs:2.65,3.8775)
--(axis cs:2.6,3.84)
--(axis cs:2.55,3.7975)
--(axis cs:2.5,3.75)
--(axis cs:2.45,3.6975)
--(axis cs:2.4,3.64)
--(axis cs:2.35,3.5775)
--(axis cs:2.3,3.51)
--(axis cs:2.25,3.4375)
--(axis cs:2.2,3.36)
--(axis cs:2.15,3.2775)
--(axis cs:2.1,3.19)
--(axis cs:2.05,3.0975)
--(axis cs:2,3)
--(axis cs:1.95,2.8975)
--(axis cs:1.9,2.79)
--(axis cs:1.85,2.6775)
--(axis cs:1.8,2.56)
--(axis cs:1.75,2.4375)
--(axis cs:1.7,2.31)
--(axis cs:1.65,2.1775)
--(axis cs:1.6,2.04)
--(axis cs:1.55,1.8975)
--(axis cs:1.5,1.75)
--(axis cs:1.45,1.5975)
--(axis cs:1.4,1.44)
--(axis cs:1.35,1.2775)
--(axis cs:1.3,1.11)
--(axis cs:1.25,0.937500000000001)
--(axis cs:1.2,0.760000000000001)
--(axis cs:1.15,0.5775)
--(axis cs:1.1,0.39)
--(axis cs:1.05,0.1975)
--(axis cs:1,0)
--cycle;

\addplot [semithick, color0]
table {%
1.1 0.39
1.1 0.375
1.1 0.36
1.1 0.345
1.1 0.33
1.1 0.315
1.1 0.3
1.1 0.285
1.1 0.27
1.1 0.255
1.1 0.24
1.1 0.225
1.1 0.21
1.1 0.195
1.1 0.18
1.1 0.165
1.1 0.15
1.1 0.135
1.1 0.12
1.1 0.105
1.1 0.0900000000000001
};
\addplot [semithick, color1]
table {%
1.4 1.44
1.4 1.38
1.4 1.32
1.4 1.26
1.4 1.2
1.4 1.14
1.4 1.08
1.4 1.02
1.4 0.96
1.4 0.9
1.4 0.84
1.4 0.78
1.4 0.72
1.4 0.66
1.4 0.6
1.4 0.54
1.4 0.48
1.4 0.42
1.4 0.36
1.4 0.3
1.4 0.24
};
\addplot [semithick, color2]
table {%
2.1 3.19
2.1 3.035
2.1 2.88
2.1 2.725
2.1 2.57
2.1 2.415
2.1 2.26
2.1 2.105
2.1 1.95
2.1 1.795
2.1 1.64
2.1 1.485
2.1 1.33
2.1 1.175
2.1 1.02
2.1 0.865000000000001
2.1 0.71
2.1 0.555000000000001
2.1 0.4
2.1 0.244999999999999
2.1 0.0899999999999999
};
\addplot [semithick, white!46.6666666666667!black]
table {%
2.6 3.84
2.6 3.66
2.6 3.48
2.6 3.3
2.6 3.12
2.6 2.94
2.6 2.76
2.6 2.58
2.6 2.4
2.6 2.22
2.6 2.04
2.6 1.86
2.6 1.68
2.6 1.5
2.6 1.32
2.6 1.14
2.6 0.959999999999999
2.6 0.779999999999999
2.6 0.599999999999999
2.6 0.419999999999999
2.6 0.239999999999999
};
\addplot [semithick, color3]
table {%
3 4
3 3.8
3 3.6
3 3.4
3 3.2
3 3
3 2.8
3 2.6
3 2.4
3 2.2
3 2
3 1.8
3 1.6
3 1.4
3 1.2
3 1
3 0.800000000000001
3 0.6
3 0.4
3 0.199999999999996
3 0
};
\addplot [semithick, color4]
table {%
3.9 3.19
3.9 3.035
3.9 2.88
3.9 2.725
3.9 2.57
3.9 2.415
3.9 2.26
3.9 2.105
3.9 1.95
3.9 1.795
3.9 1.64
3.9 1.485
3.9 1.33
3.9 1.175
3.9 1.02
3.9 0.865000000000004
3.9 0.709999999999997
3.9 0.554999999999998
3.9 0.4
3.9 0.245000000000001
3.9 0.0899999999999999
};
\addplot [semithick, color5]
table {%
4.7 1.11
4.7 1.065
4.7 1.02
4.7 0.975000000000001
4.7 0.929999999999993
4.7 0.884999999999994
4.7 0.839999999999996
4.7 0.794999999999998
4.7 0.749999999999993
4.7 0.704999999999995
4.7 0.659999999999997
4.7 0.614999999999998
4.7 0.569999999999993
4.7 0.524999999999995
4.7 0.480000000000004
4.7 0.434999999999999
4.7 0.389999999999997
4.7 0.345000000000006
4.7 0.300000000000004
4.7 0.254999999999995
4.7 0.209999999999997
};
\end{axis}

\end{tikzpicture}}%
        \label{fig:ghost_gsd_b}}
    \\
    \subfloat[]{\resizebox{0.23\textwidth}{!}{\input{plots/space_sigma.tex}}%
        \label{fig:ghost_ordered_probit_c}}
    \hfil
    \subfloat[]{\resizebox{0.23\textwidth}{!}{\input{plots/ghost_sigma.tex}}%
        \label{fig:ghost_ordered_probit_d}}
    \hfil
    \subfloat[]{\resizebox{0.23\textwidth}{!}{
\begin{tikzpicture}

\definecolor{color0}{rgb}{0.886274509803922,0.290196078431373,0.2}
\definecolor{color1}{rgb}{0.203921568627451,0.541176470588235,0.741176470588235}
\definecolor{color2}{rgb}{0.596078431372549,0.556862745098039,0.835294117647059}
\definecolor{color3}{rgb}{0.984313725490196,0.756862745098039,0.368627450980392}
\definecolor{color4}{rgb}{0.556862745098039,0.729411764705882,0.258823529411765}
\definecolor{color5}{rgb}{1,0.709803921568627,0.72156862745098}

\begin{axis}[
axis background/.style={fill=white},
axis line style={white!89.8039215686275!black},
tick align=outside,
tick pos=left,
x grid style={white!89.8039215686275!black},
xlabel={\(\displaystyle \psi\)},
xmajorgrids,
xmin=0.8, xmax=5.2,
xtick style={color=white!33.3333333333333!black},
y grid style={white!89.8039215686275!black},
ylabel={\(\displaystyle \rho\)},
ymajorgrids,
ymin=-0.0828125, ymax=1.0515625,
ytick style={color=white!33.3333333333333!black},
ytick={-0.2,0,0.2,0.4,0.6,0.8,1,1.2},
yticklabels={\ensuremath{-}0.2,0.0,0.2,0.4,0.6,0.8,1.0,1.2}
]
\path [draw=blue, fill=blue, opacity=0.15, very thin]
(axis cs:1,0)
--(axis cs:1,-0.03125)
--(axis cs:5,-0.03125)
--(axis cs:5,0)
--(axis cs:5,0)
--(axis cs:1,0)
--cycle;

\addplot [semithick, color0]
table {%
1.01 0.01
1.06 0.01
1.11 0.01
1.16 0.01
1.21 0.01
1.26 0.01
1.31 0.01
1.36 0.01
1.41 0.01
1.46 0.01
1.51 0.01
1.56 0.01
1.61 0.01
1.66 0.01
1.71 0.01
1.76 0.01
1.81 0.01
1.86 0.01
1.91 0.01
1.96 0.01
2.01 0.01
2.06 0.01
2.11 0.01
2.16 0.01
2.21 0.01
2.26 0.01
2.31 0.01
2.36 0.01
2.41 0.01
2.46 0.01
2.51 0.01
2.56 0.01
2.61 0.01
2.66 0.01
2.71 0.01
2.76 0.01
2.81 0.01
2.86 0.01
2.91 0.01
2.96 0.01
3.01 0.01
3.06 0.01
3.11 0.01
3.16 0.01
3.21 0.01
3.26 0.01
3.31 0.01
3.36 0.01
3.41 0.01
3.46 0.01
3.51 0.01
3.56 0.01
3.61 0.01
3.66 0.01
3.71 0.01
3.76 0.01
3.81 0.01
3.86 0.01
3.91 0.01
3.96 0.01
4.01 0.01
4.06 0.01
4.11 0.01
4.16 0.01
4.21 0.01
4.26 0.01
4.31 0.01
4.36 0.01
4.41 0.01
4.46 0.01
4.51 0.01
4.56 0.01
4.61 0.01
4.66 0.01
4.71 0.01
4.76 0.01
4.81 0.01
4.86 0.01
4.91 0.01
4.96 0.01
4.99 0.01
};
\addplot [semithick, color1]
table {%
1.01 0.1
1.06 0.1
1.11 0.1
1.16 0.1
1.21 0.1
1.26 0.1
1.31 0.1
1.36 0.1
1.41 0.1
1.46 0.1
1.51 0.1
1.56 0.1
1.61 0.1
1.66 0.1
1.71 0.1
1.76 0.1
1.81 0.1
1.86 0.1
1.91 0.1
1.96 0.1
2.01 0.1
2.06 0.1
2.11 0.1
2.16 0.1
2.21 0.1
2.26 0.1
2.31 0.1
2.36 0.1
2.41 0.1
2.46 0.1
2.51 0.1
2.56 0.1
2.61 0.1
2.66 0.1
2.71 0.1
2.76 0.1
2.81 0.1
2.86 0.1
2.91 0.1
2.96 0.1
3.01 0.1
3.06 0.1
3.11 0.1
3.16 0.1
3.21 0.1
3.26 0.1
3.31 0.1
3.36 0.1
3.41 0.1
3.46 0.1
3.51 0.1
3.56 0.1
3.61 0.1
3.66 0.1
3.71 0.1
3.76 0.1
3.81 0.1
3.86 0.1
3.91 0.1
3.96 0.1
4.01 0.1
4.06 0.1
4.11 0.1
4.16 0.1
4.21 0.1
4.26 0.1
4.31 0.1
4.36 0.1
4.41 0.1
4.46 0.1
4.51 0.1
4.56 0.1
4.61 0.1
4.66 0.1
4.71 0.1
4.76 0.1
4.81 0.1
4.86 0.1
4.91 0.1
4.96 0.1
4.99 0.1
};
\addplot [semithick, color2]
table {%
1.01 0.4
1.06 0.4
1.11 0.4
1.16 0.4
1.21 0.4
1.26 0.4
1.31 0.4
1.36 0.4
1.41 0.4
1.46 0.4
1.51 0.4
1.56 0.4
1.61 0.4
1.66 0.4
1.71 0.4
1.76 0.4
1.81 0.4
1.86 0.4
1.91 0.4
1.96 0.4
2.01 0.4
2.06 0.4
2.11 0.4
2.16 0.4
2.21 0.4
2.26 0.4
2.31 0.4
2.36 0.4
2.41 0.4
2.46 0.4
2.51 0.4
2.56 0.4
2.61 0.4
2.66 0.4
2.71 0.4
2.76 0.4
2.81 0.4
2.86 0.4
2.91 0.4
2.96 0.4
3.01 0.4
3.06 0.4
3.11 0.4
3.16 0.4
3.21 0.4
3.26 0.4
3.31 0.4
3.36 0.4
3.41 0.4
3.46 0.4
3.51 0.4
3.56 0.4
3.61 0.4
3.66 0.4
3.71 0.4
3.76 0.4
3.81 0.4
3.86 0.4
3.91 0.4
3.96 0.4
4.01 0.4
4.06 0.4
4.11 0.4
4.16 0.4
4.21 0.4
4.26 0.4
4.31 0.4
4.36 0.4
4.41 0.4
4.46 0.4
4.51 0.4
4.56 0.4
4.61 0.4
4.66 0.4
4.71 0.4
4.76 0.4
4.81 0.4
4.86 0.4
4.91 0.4
4.96 0.4
4.99 0.4
};
\addplot [semithick, white!46.6666666666667!black]
table {%
1.01 0.6
1.06 0.6
1.11 0.6
1.16 0.6
1.21 0.6
1.26 0.6
1.31 0.6
1.36 0.6
1.41 0.6
1.46 0.6
1.51 0.6
1.56 0.6
1.61 0.6
1.66 0.6
1.71 0.6
1.76 0.6
1.81 0.6
1.86 0.6
1.91 0.6
1.96 0.6
2.01 0.6
2.06 0.6
2.11 0.6
2.16 0.6
2.21 0.6
2.26 0.6
2.31 0.6
2.36 0.6
2.41 0.6
2.46 0.6
2.51 0.6
2.56 0.6
2.61 0.6
2.66 0.6
2.71 0.6
2.76 0.6
2.81 0.6
2.86 0.6
2.91 0.6
2.96 0.6
3.01 0.6
3.06 0.6
3.11 0.6
3.16 0.6
3.21 0.6
3.26 0.6
3.31 0.6
3.36 0.6
3.41 0.6
3.46 0.6
3.51 0.6
3.56 0.6
3.61 0.6
3.66 0.6
3.71 0.6
3.76 0.6
3.81 0.6
3.86 0.6
3.91 0.6
3.96 0.6
4.01 0.6
4.06 0.6
4.11 0.6
4.16 0.6
4.21 0.6
4.26 0.6
4.31 0.6
4.36 0.6
4.41 0.6
4.46 0.6
4.51 0.6
4.56 0.6
4.61 0.6
4.66 0.6
4.71 0.6
4.76 0.6
4.81 0.6
4.86 0.6
4.91 0.6
4.96 0.6
4.99 0.6
};
\addplot [semithick, color3]
table {%
1.01 0.85
1.06 0.85
1.11 0.85
1.16 0.85
1.21 0.85
1.26 0.85
1.31 0.85
1.36 0.85
1.41 0.85
1.46 0.85
1.51 0.85
1.56 0.85
1.61 0.85
1.66 0.85
1.71 0.85
1.76 0.85
1.81 0.85
1.86 0.85
1.91 0.85
1.96 0.85
2.01 0.85
2.06 0.85
2.11 0.85
2.16 0.85
2.21 0.85
2.26 0.85
2.31 0.85
2.36 0.85
2.41 0.85
2.46 0.85
2.51 0.85
2.56 0.85
2.61 0.85
2.66 0.85
2.71 0.85
2.76 0.85
2.81 0.85
2.86 0.85
2.91 0.85
2.96 0.85
3.01 0.85
3.06 0.85
3.11 0.85
3.16 0.85
3.21 0.85
3.26 0.85
3.31 0.85
3.36 0.85
3.41 0.85
3.46 0.85
3.51 0.85
3.56 0.85
3.61 0.85
3.66 0.85
3.71 0.85
3.76 0.85
3.81 0.85
3.86 0.85
3.91 0.85
3.96 0.85
4.01 0.85
4.06 0.85
4.11 0.85
4.16 0.85
4.21 0.85
4.26 0.85
4.31 0.85
4.36 0.85
4.41 0.85
4.46 0.85
4.51 0.85
4.56 0.85
4.61 0.85
4.66 0.85
4.71 0.85
4.76 0.85
4.81 0.85
4.86 0.85
4.91 0.85
4.96 0.85
4.99 0.85
};
\addplot [semithick, color4]
table {%
1.01 0.92
1.06 0.92
1.11 0.92
1.16 0.92
1.21 0.92
1.26 0.92
1.31 0.92
1.36 0.92
1.41 0.92
1.46 0.92
1.51 0.92
1.56 0.92
1.61 0.92
1.66 0.92
1.71 0.92
1.76 0.92
1.81 0.92
1.86 0.92
1.91 0.92
1.96 0.92
2.01 0.92
2.06 0.92
2.11 0.92
2.16 0.92
2.21 0.92
2.26 0.92
2.31 0.92
2.36 0.92
2.41 0.92
2.46 0.92
2.51 0.92
2.56 0.92
2.61 0.92
2.66 0.92
2.71 0.92
2.76 0.92
2.81 0.92
2.86 0.92
2.91 0.92
2.96 0.92
3.01 0.92
3.06 0.92
3.11 0.92
3.16 0.92
3.21 0.92
3.26 0.92
3.31 0.92
3.36 0.92
3.41 0.92
3.46 0.92
3.51 0.92
3.56 0.92
3.61 0.92
3.66 0.92
3.71 0.92
3.76 0.92
3.81 0.92
3.86 0.92
3.91 0.92
3.96 0.92
4.01 0.92
4.06 0.92
4.11 0.92
4.16 0.92
4.21 0.92
4.26 0.92
4.31 0.92
4.36 0.92
4.41 0.92
4.46 0.92
4.51 0.92
4.56 0.92
4.61 0.92
4.66 0.92
4.71 0.92
4.76 0.92
4.81 0.92
4.86 0.92
4.91 0.92
4.96 0.92
4.99 0.92
};
\addplot [semithick, color5]
table {%
1.01 1
1.06 1
1.11 1
1.16 1
1.21 1
1.26 1
1.31 1
1.36 1
1.41 1
1.46 1
1.51 1
1.56 1
1.61 1
1.66 1
1.71 1
1.76 1
1.81 1
1.86 1
1.91 1
1.96 1
2.01 1
2.06 1
2.11 1
2.16 1
2.21 1
2.26 1
2.31 1
2.36 1
2.41 1
2.46 1
2.51 1
2.56 1
2.61 1
2.66 1
2.71 1
2.76 1
2.81 1
2.86 1
2.91 1
2.96 1
3.01 1
3.06 1
3.11 1
3.16 1
3.21 1
3.26 1
3.31 1
3.36 1
3.41 1
3.46 1
3.51 1
3.56 1
3.61 1
3.66 1
3.71 1
3.76 1
3.81 1
3.86 1
3.91 1
3.96 1
4.01 1
4.06 1
4.11 1
4.16 1
4.21 1
4.26 1
4.31 1
4.36 1
4.41 1
4.46 1
4.51 1
4.56 1
4.61 1
4.66 1
4.71 1
4.76 1
4.81 1
4.86 1
4.91 1
4.96 1
4.99 1
};
\end{axis}

\end{tikzpicture}}%
        \label{fig:ghost_gsd_c}}
    \hfil
    \subfloat[]{\resizebox{0.23\textwidth}{!}{\input{plots/ghost_rho.tex}}%
        \label{fig:ghost_gsd_d}}
    \caption{Mapping of Ordered Probit parameters to the $E(U)$
    and $V(U)$ space (plots (a), (b), (e) and (f)). Mapping of
    GSD parameters to the $E(U)$ and $V(U)$ space (plots (c), (d), (g)
    and (h)).
    The violet area marks all possible ($E(U), V(U)$)
    pairs for a discrete process with values $\{1, 2, 3, 4, 5\}$. The
    violet bar below plots (a), (c), (e) and (g) shows the 1–5 range (reflecting
    the range of values of random variable $U$).}
    \label{fig:ghost_figures}
\end{figure*}

Fig.~\ref{fig:ghost_figures} presents how Ordered Probit model parameters
map to the $E(U)$ and $V(U)$ space. In other words, the figure shows how parameters
of the Ordered Probit model we use to describe observed data 
(cf. Fig.~\ref{fig:ghost_ordered_probit_a} and Fig.~\ref{fig:ghost_ordered_probit_c})
map to summary statistics
computed directly on these observed data (Fig.~\ref{fig:ghost_ordered_probit_b} and
Fig.~\ref{fig:ghost_ordered_probit_d}).
Intuitively, Fig.~\ref{fig:ghost_ordered_probit_a} and
Fig.~\ref{fig:ghost_ordered_probit_c}
show us
how the Ordered Probit model ``sees'' observed data.
Fig.~\ref{fig:ghost_ordered_probit_b} and
Fig.~\ref{fig:ghost_ordered_probit_d} show
us how observed data actually look like in terms of two basic summary statistics (i.e.,
mean $E(U)$ and variance $V(U)$).
Differently put, any point along any line in
Fig.~\ref{fig:ghost_ordered_probit_a} or
Fig.~\ref{fig:ghost_ordered_probit_c} corresponds to a fixed pair of
Ordered Probit parameters. The Ordered Probit model with
these parameters is then used to generate discrete responses (being
realisations of random variable $U$). Summary statistics
(i.e., $E(U)$ and $V(U)$) computed on
these generated responses yield a single point in
Fig.~\ref{fig:ghost_ordered_probit_b} or
Fig.~\ref{fig:ghost_ordered_probit_d}, respectively.
(Note that these generated responses can be thought of as
representing individual responses
we observe in real subjective experiments.)
Importantly, plots in Fig.~\ref{fig:ghost_figures}
should be analysed in pairs, row-wise. Stated differently,
the leftmost (red) line in Fig.~\ref{fig:ghost_ordered_probit_a} corresponds to the same data
series as the leftmost (red) line in Fig.~\ref{fig:ghost_ordered_probit_b}. The same is true
for Fig.~\ref{fig:ghost_ordered_probit_c} and Fig.~\ref{fig:ghost_ordered_probit_d}, and
so on.
When analysing Fig.~\ref{fig:ghost_figures}, please also
keep in mind that $E(O)=\mu$ and $V(O)=\sigma^2$ (cf.
Sec.~\ref{ssec:subjective_response_as_a_random_variable} for more context).
%
%

We want model parameters to accurately reflect phenomena
occurring in observed data. For example, we naturally associate the $\mu$ parameter
with the central tendency of observed data and the $\sigma$ parameter with their
variance. Thus, if we keep $\mu$ constant and
increase the value of $\sigma$, we expect this should correspond to $E(U)$ staying
constant and $V(U)$ to increase. However, this is not the case.
Instead, keeping $\mu$ constant and increasing $\sigma$ corresponds to
changes in both $E(U)$ and $V(U)$.
This can be observed by following same-coloured
lines\footnote{The ordering of lines in Fig.~\ref{fig:ghost_ordered_probit_a}
and Fig.~\ref{fig:ghost_ordered_probit_c} is the same as the ordering
of lines in Fig.~\ref{fig:ghost_ordered_probit_b} and
Fig.~\ref{fig:ghost_ordered_probit_d}. Thus, the figure can be interpreted
in black-and-white print as well.}
in
Fig.~\ref{fig:ghost_ordered_probit_a} and Fig.~\ref{fig:ghost_ordered_probit_b}.
Specifically, let us take the leftmost (red) line from
Fig.~\ref{fig:ghost_ordered_probit_a}. It corresponds to Ordered Probit's
$\mu$ fixed at a value slightly larger than zero. Moving vertically
upwards along this line, $\mu$ stays constant and $\sigma$ increases. If
we were to stop at various points along this line and generate discrete
responses (being realisations of random variable $U$)
from the Ordered Probit model with $\mu$ and $\sigma$ parameters
fixed, we expect each such sample should have a constant and same
expected value $E(U)$, but changing variance $V(U)$. The corresponding
leftmost (red) line in Fig.~\ref{fig:ghost_ordered_probit_b} shows
what $E(U)$ and $V(U)$ we actually observe when generating responses
from the Ordered Probit model. As we can see, the samples generated do
not have constant expected value. On the contrary, it changes in rather
unexpected way, as we move along increasing values of $\sigma$.
(The only exception to this rule is when $\mu=3$.) This property of Ordered Probit
model parameters makes them counter-intuitive.
Unfortunately, this is not the only
limitation of Ordered Probit's parameterisation. Another one relates to how changes in
$\mu$ correspond to changes in $E(U)$. Looking at Fig.~\ref{fig:ghost_ordered_probit_c}
and Fig.~\ref{fig:ghost_ordered_probit_d} we see that the same range of $\mu$ values
maps to different ranges of $E(U)$ values as the $\sigma$ parameter changes. For example,
let us compare the topmost pink curve ($\sigma = 8$)
with the second topmost green one ($2 < \sigma < 4$). In
Fig.~\ref{fig:ghost_ordered_probit_c} they both span the same range of $\mu$ values.
However, in Fig.~\ref{fig:ghost_ordered_probit_d}, the pink curve corresponds to much
narrower range of $E(U)$, when compared to the green curve. This leads us to another
limitation of Ordered Probit's parameterisation. Although subjective responses we
take into account here span the range from 1 to 5, the $\mu$ parameter takes values
exceeding the 1–5 range. Practically speaking, although it is tempting to treat
$\mu$ as an MOS-related measure, $\mu$ can and will
exceed the 1–5 range (which the MOS never does).
Thus, $\mu$ should not be intuitively interpreted as MOS
counterpart for the Ordered Probit model. For completeness, we mention that
the Ordered Probit model is able to describe the complete ghost-like area
shown in Fig.~\ref{fig:ghost_ordered_probit_b} and
Fig.~\ref{fig:ghost_ordered_probit_d}. However, this is only possible if
we allow its parameters to change without bounds. In other words,
when $(\mu, \sigma) \in (-\infty, +\infty) \times (0, +\infty)$.

The right hand side of Fig.~\ref{fig:ghost_figures}
is GSD's counterpart of Ordered Probit plots on the left.
Fig.~\ref{fig:ghost_gsd_a} and Fig.~\ref{fig:ghost_gsd_c}
present GSD parameters space. Fig.~\ref{fig:ghost_gsd_b}
and Fig.~\ref{fig:ghost_gsd_d}
present the $E(U)$ and $V(U)$ space. (Note that there is an inverse relationship
between $\rho$ and $V(U)$.)
As can be readily seen, the GSD does not
suffer from problems inherent to the Ordered Probit model. In particular,
keeping the $\psi$ parameter constant and changing $\rho$ parameter's value, we keep
the $E(U)$ constant and vary $V(U)$ only. This means GSD's parameterisation allows
for treating $\psi$ as observed data's central tendency and $\rho$ as a measure of
their variability. Following same-coloured lines in Fig.~\ref{fig:ghost_gsd_a} and
Fig.~\ref{fig:ghost_gsd_b} evidences how keeping $\psi$ constant corresponds to
constant $E(U)$. Please also notice that the same range of $\psi$ values for different
values of $\rho$, always corresponds to the same range of $E(U)$. We
can take a curve of any colour from Fig.~\ref{fig:ghost_gsd_c} and
Fig.~\ref{fig:ghost_gsd_d}, and see that it always spans the entire range of $E(U)$.
Although the bumpy shape of multiple curves in Fig.~\ref{fig:ghost_gsd_d} may seem
counter-intuitive at first, it reflects an important property of $\rho$. The $\rho$
parameter expresses what ratio of available variance for a given mean is present in
observed data. Thus, to keep this ratio constant across different means, the curve has
to follow the bottom part of the $E(U)$ and $V(U)$ space.
Thanks to its properties, $\rho=0.5$ means that we deal with data being
at the midpoint between minimum and maximum possible variance.
Finally, GSD parameters cover the entire space of $E(U)$ and $V(U)$, and do so
staying in the well defined bounds. Specifically, $(\psi, \rho) \in [1, 5] \times [0, 1]$.
Practically speaking, $\psi$ can be regarded as GSD's counterpart of the MOS.

We should note here that both models share one limitation. Even when data
variability related parameter ($\sigma$ or $\rho$) stays constant and central
tendency related parameter ($\mu$ or $\psi$) changes, $V(U)$ changes across
different values of $E(U)$. Ideally, $V(U)$ should follow variability related
parameter and stay constant across chaning $E(U)$.
However, since we are dealing here with discrete,
limited domain process (only values $\{1, 2, 3, 4, 5\}$ can be observed), the
mean is naturally coupled with variance. In other words, changes to the mean
inherently influence variance.

\subsection{Good Description of Typical MQA Experiments}
\label{ssec:good_description_of_typical_mqa_experiments}
\begin{figure*}[!t]
    \centering
    \subfloat[]{\resizebox{.3\textwidth}{!}{\includegraphics[width=3.49in]{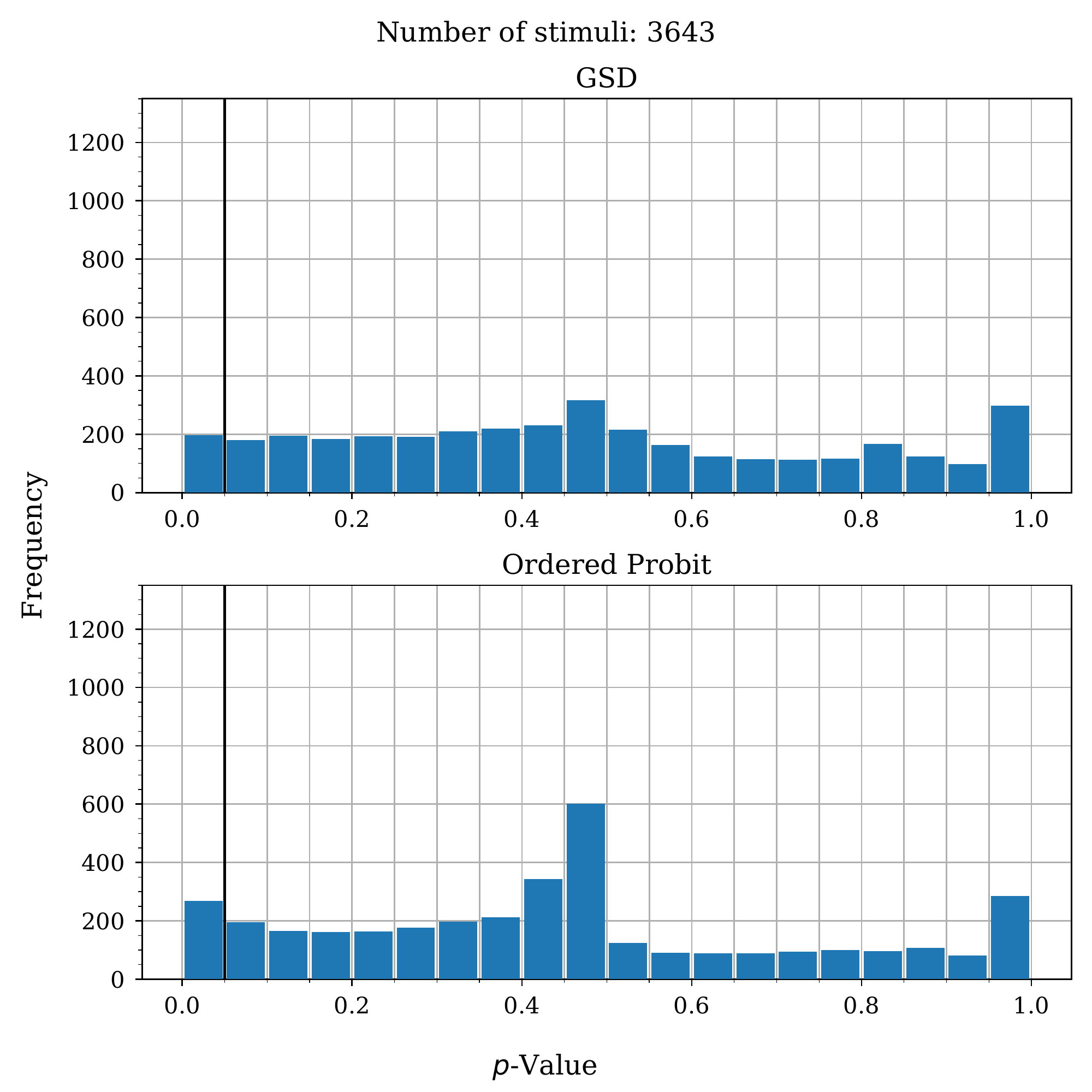}}%
        \label{fig:p_value_histogram_mqa_exp}}
    \hfil
    \subfloat[]{\resizebox{.3\textwidth}{!}{\includegraphics[width=3.49in]{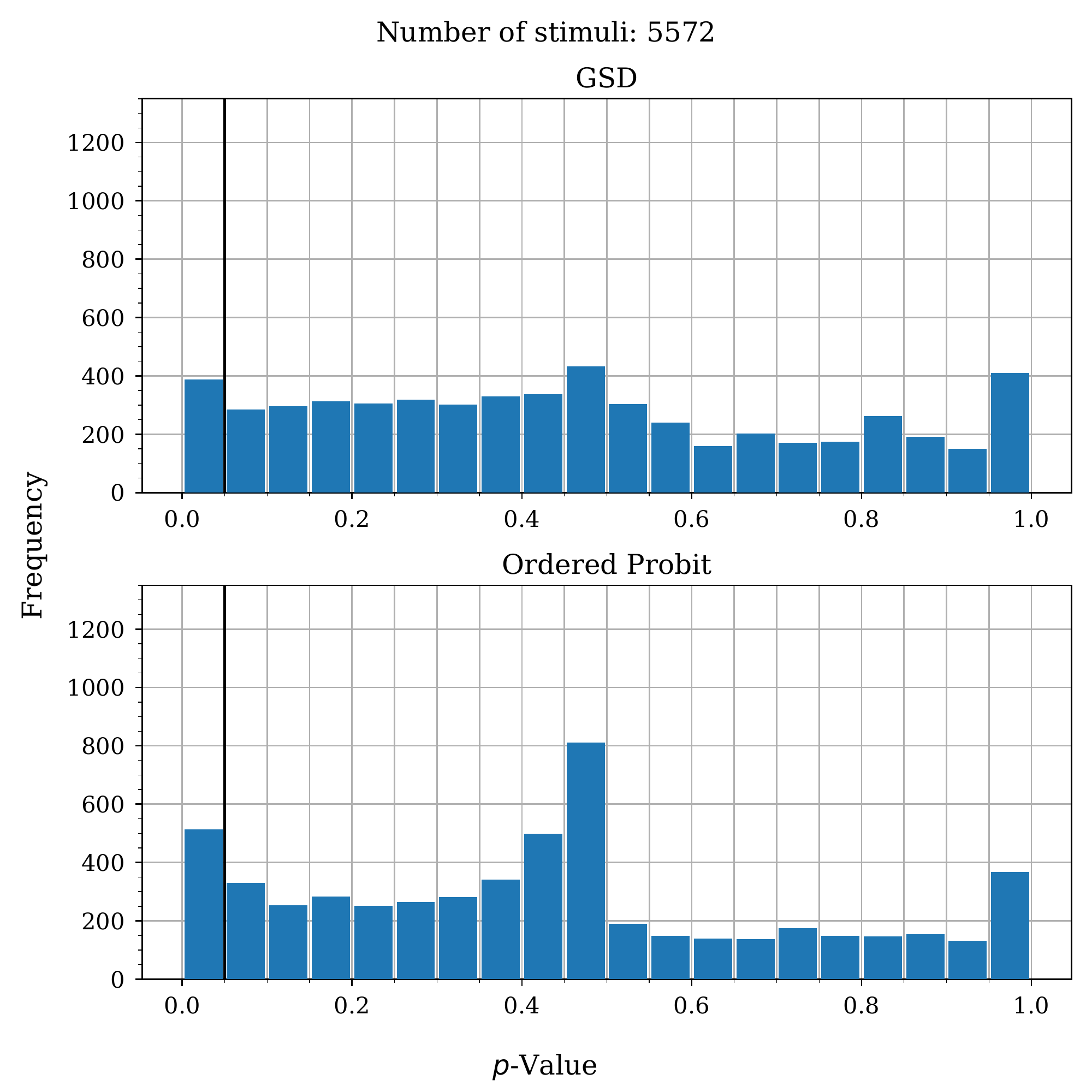}}%
        \label{fig:p_value_histogram_broad_mqa_exp}}
    \hfil
    \subfloat[]{\resizebox{.3\textwidth}{!}{\includegraphics[width=3.49in]{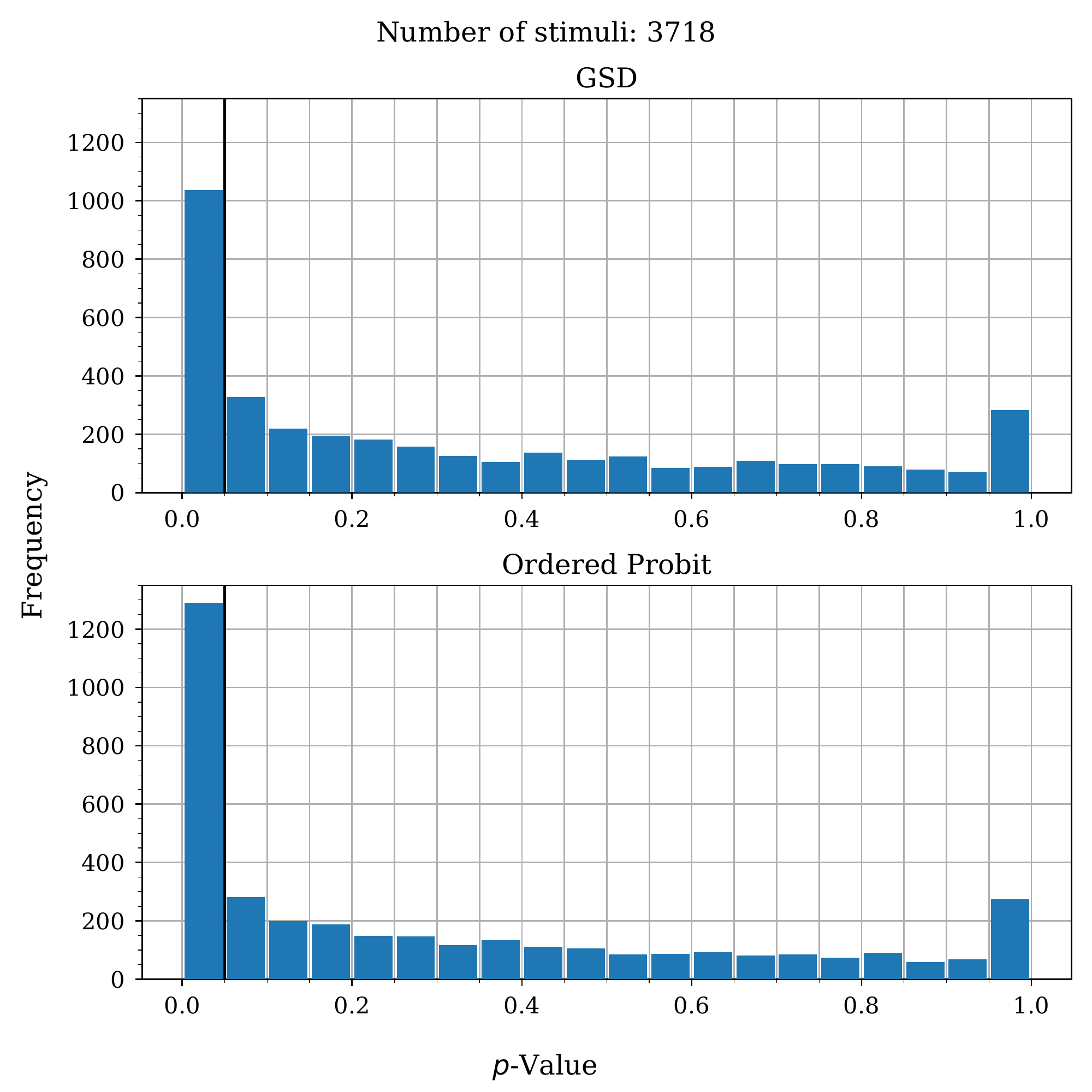}}%
        \label{fig:p_value_histogram_non_mqa_exp}}
    \caption{$p$-Value histograms for the GSD (upper) and Ordered Probit (lower) models.
	$p$-Values come from the G-test of goodness-of-fit applied to stimuli from
	(a) typical Multimedia Quality Assessment (MQA) experiments,
	(b) typical and broadly understood MQA experiments and
	(c) non-MQA experiments.
	The thick vertical line marks the 0.05 significance level.}
    \label{fig:p_value_histograms}
\end{figure*}

\begin{figure}[!t]
    \centering
    \includegraphics[width=3.49in, viewport=12 10 340 195, clip]{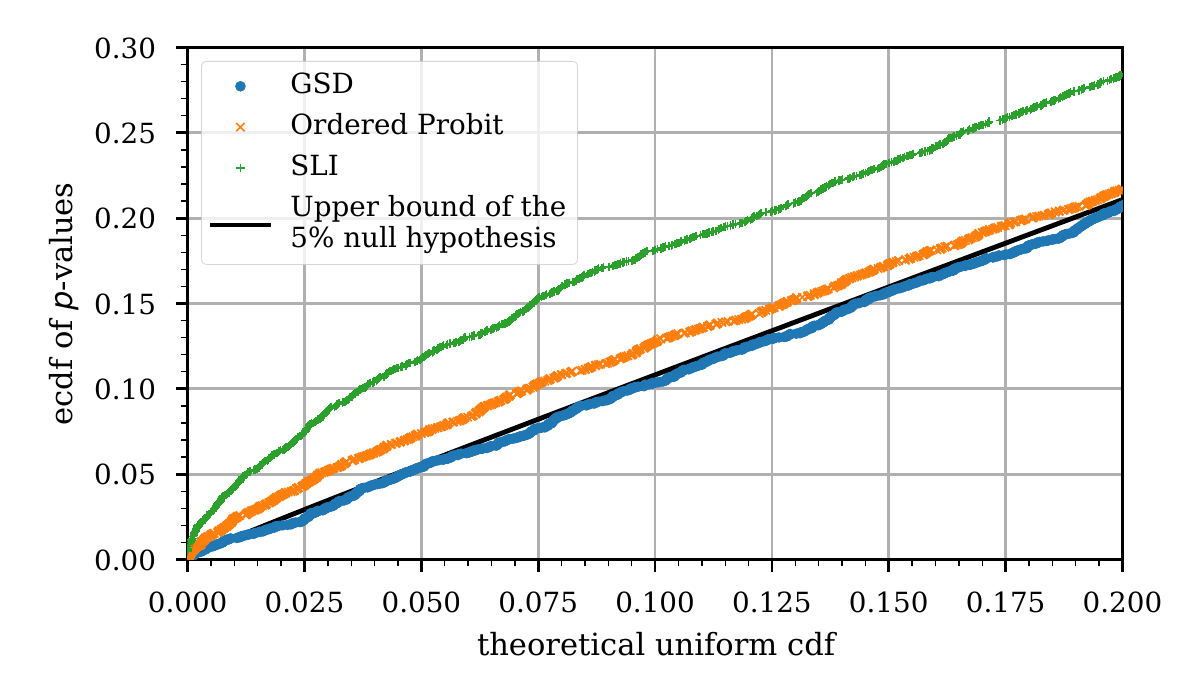}
    \caption{$p$-Value P–P plot for typical MQA
	experiments. $p$-Values come from the G-test of goodness-of-fit applied to the
	GSD, Ordered Probit and Simplified Li2020 (SLI) models, fitted to responses from typical MQA
	experiments. CDF stands
	for cumulative distribution function and ECDF for empirical cumulative distribution function.}
    \label{fig:pvalue_pp_plot_mqa_exp}
\end{figure}

Fig.~\ref{fig:p_value_histogram_mqa_exp} shows results of applying a bootstrapped G-test of goodness of fit
to responses from typical MQA experiments, as modelled by the GSD or by the Ordered Probit model.
If any of the two models truly reflects response distributions observed in real data, a related
\textit{p}-value histogram should resemble the uniform distribution (or any other nonincreasing
distribution) in the region of low $p$-values (roughly between 0 and 0.2) \cite{Nawala2020ACM}.
It is easy to see that the histogram for the Ordered Probit model does not resemble
the uniform distribution. The most important indication of this fact is the height of the leftmost
bar, which is significantly greater than that of the second leftmost bar.
GSD's histogram does resemble the uniform distribution for the $p$-values region of interest.
However, to decisively assess GSD's
performance we need to resort to \textit{p}-value P–P plot
(cf. Fig.~\ref{fig:pvalue_pp_plot_mqa_exp}).
Since all GSD-related data points fall below the black diagonal line, we can safely say that results
do not contradict the null hypothesis of the GSD truly reflecting response distributions observed
in real data. In other words, the GSD well reflects response distributions observed in typical
MQA experiments. The same is not true for the Ordered Probit model. This is indicated by
all Ordered Probit related data points falling above the black diagonal line. Differently put,
the Ordered Probit model is not properly reflecting response distributions observed in typical
MQA experiments.

If we now also take into account MQA experiments that do not strictly follow international
recommendations (let us call them \textit{broadly understood MQA experiments}),
we see that performance of the both models deteriorates
(cf. Fig.~\ref{fig:p_value_histogram_broad_mqa_exp}). The best indication of this is the height
of the leftmost bar. On both histograms its height is significantly greater than the reference
height corresponding to approximately 279 stimuli or 5\% of all stimuli investigated. We do not
show a related P–P plot since it simply reaffirms both models do not reflect response distributions
observed in real data.

We also investigated how the GSD and Ordered Probit models would perform on a data set not
related to MQA experiments. For this we chose two data sets popular in the movie recommendation
systems research community: (i) MovieLens 1M~\cite{Harper2015} and (ii)
Personality 2018~\cite{Nguyen2018}. Although the two data sets are outside of MQA, they use
the 5-level Likert scale to collect subjective responses. Our hypothesis was that since the
GSD performed well on MQA data using the 5-level Liker scale, then maybe it would perform well
on these data sets as well. However, looking at Fig.~\ref{fig:p_value_histogram_non_mqa_exp},
we can clearly see that both the GSD and Ordered Probit models do not reflect response distributions
observed in the data. In other words, neither the GSD nor Ordered Probit model well describe
response distributions observed in data concerning movie recommendation systems.

\subsection{Better Than Empirical Distribution}
\label{ssec:better_than_empirical_distribution}
It is interesting to check whether the GSD brings any advantage if it
comes to generalisability. We define generalisability as the ability
of a model to capture large sample phenomena when observing only a
subsample of the large sample. In particular, we would like to check
whether the GSD better captures large sample's distribution shape in
comparison to the empirical distribution of the subsample. Put differently,
we would like to check whether the GSD is better suited
for bootstrapping than is the empirical distribution. If this
proves to be the case, then the GSD could be used for resampling.
One important consequence of this would be a chance
to build better machine learning (ML) models
aimed at predicting subjectively perceived multimedia quality
(which is a difficult, important and still open challenge).
It is often the case in the
field of Multimedia Quality Assessment (MQA) that only up to 30 responses
per stimulus are available. If one wants to create an ML model this
may prove insufficient and resampling must be applied to generate more
responses per stimulus. Should the GSD prove to be better
for bootstrapping than the empirical distribution
(which is typically applied in this context), the GSD could be used to
generate more reliable samples during resampling.

To test GSD's generalisability capabilities in practice we
use data from four MQA studies: (i) MM2~\cite{Pinson2012},
(ii) VQEG HDTV Phase I~\cite{HDTV_Phase_I_test},
(iii) NFLX (cf. Section~\ref{ssec:data_sets} to learn more about this study)
and (iv) ITERO~\cite{Perez2021}.
From these studies
we extract responses for selected stimuli. More precisely, we select
stimuli with at least 144 responses. This way we get 234 stimuli.
The number of responses per stimulus spans from 144 to
228. There are only four unique numbers of responses per stimulus.
Table~\ref{tab:no_of_res_per_stimulus} shows the four numbers of responses
and the count of stimuli with a given number of responses. Furthermore,
it shows from which study a given set of stimuli was taken.
\begin{table}[!t]
    \renewcommand{\arraystretch}{1.3}
    \caption{Distribution of Responses Among the Four Studies Used
    in the Bootstrap Analysis. HDTV Corresponds to
    VQEG HDTV Phase I Study.}
    \label{tab:no_of_res_per_stimulus}
    \centering
    \begin{tabular}{ccc}
    \hline\hline
    No. of Responses & No. of Stimuli & Study \\
    \hline
    144                 & 24    & HDTV          \\
    200                 & 40    & NFLX          \\
    213                 & 60    & MM2           \\
    228                 & 110   & ITERO         \\
    \hline\hline
    \end{tabular}
\end{table}

The NFLX study contains responses given to stimuli displayed on
one of the two display devices---tablet and TV.
In principle, responses for different display devices shall be analysed separately.
However, since the responses for the two devices are highly correlated
and since the same visual content was presented to participants
during the sessions with each device, we decide to
include in this analysis the combined responses from the two display devices.

If it comes to the HDTV Phase I study we only focus on responses
provided to the so called \textit{common set} of stimuli. The stimuli
from the common set were presented to participants in all the six
experiments that were part of the HDTV Phase I study. Although the
six experiments were conducted by different research teams and using
different display devices, the experimenters declared that
actions were taken to make the six experiments similar to each other.
Specifically, all video stimuli were displayed with the same
resolution and in a room conforming to guidelines of Rec. ITU-R
BT.500-11. Following experimenters declaration we combine responses
for the common set stimuli. That is, we treat the six experiments,
with 24 participants each, as one large experiment with 144 participants.
This way we end up with 24 stimuli (that many are in the common set),
each having 144 responses.

The MM2 study is a set of ten experiments. Responses in the experiments
were collected by six laboratory teams from four countries. 
Different subject pools and environments were used in each experiment.
The common
denominator of all the experiments was the same set of 60 audiovisual stimuli
and very similar test procedure. According to the authors of~\cite{Pinson2012}
the experiments were highly repeatable. Thus, we combine responses
from the ten experiments. This yields 213 responses (that many participants
in total took part in the ten experiments) for each of the 60
audiovisual stimuli.

The ITERO study collected responses from 27 subjects, who rated
the same set of 110 stimuli. The study was carried out by three
research teams. The experiment design was atypical of how MQA
experiments are usually carried out. Subjects were instructed
to repeat the experiment ten times. In total, 110 stimuli
were assigned 228 responses each (not all subjects repeated the
experiment ten times). Although the subjects were allowed
to repeat the experiment at their leisure and majority did not use
the same display device, we combine the responses from the ten
repetitions. In other words, we treat the responses as though
they come from one large subjective experiment with 110 stimuli and
228 subjects (in which each subject rates the same set of 110 stimuli).

We use three small sample sizes, i.e., $n = \{ 12, 24, 50 \}$.
This way we can observe how the GSD performs (when compared to the
empirical distribution) for different fractions of the large sample
information available. Intuitively, we expect the empirical distribution's
performance to improve as the small sample size increases. If the
GSD proofs to perform differently than the empirical distribution
we would observe how the increasing small sample size influences
the difference between the two approaches. We emphasise here that
the increasing small sample size always favours the empirical distribution.
On the other hand, the performance of the GSD depends on how well
it fits to the distribution of responses observed in the large sample.
If the fit is good, the increasing small sample size also favours
the GSD. If the fit is poor, the increasing small sample size does
not necessarily improves GSD's performance.

\begin{figure*}[!t]
    \centering
    \subfloat[]{\resizebox{.45\textwidth}{!}{\includegraphics[viewport=10 12 300 278, clip]{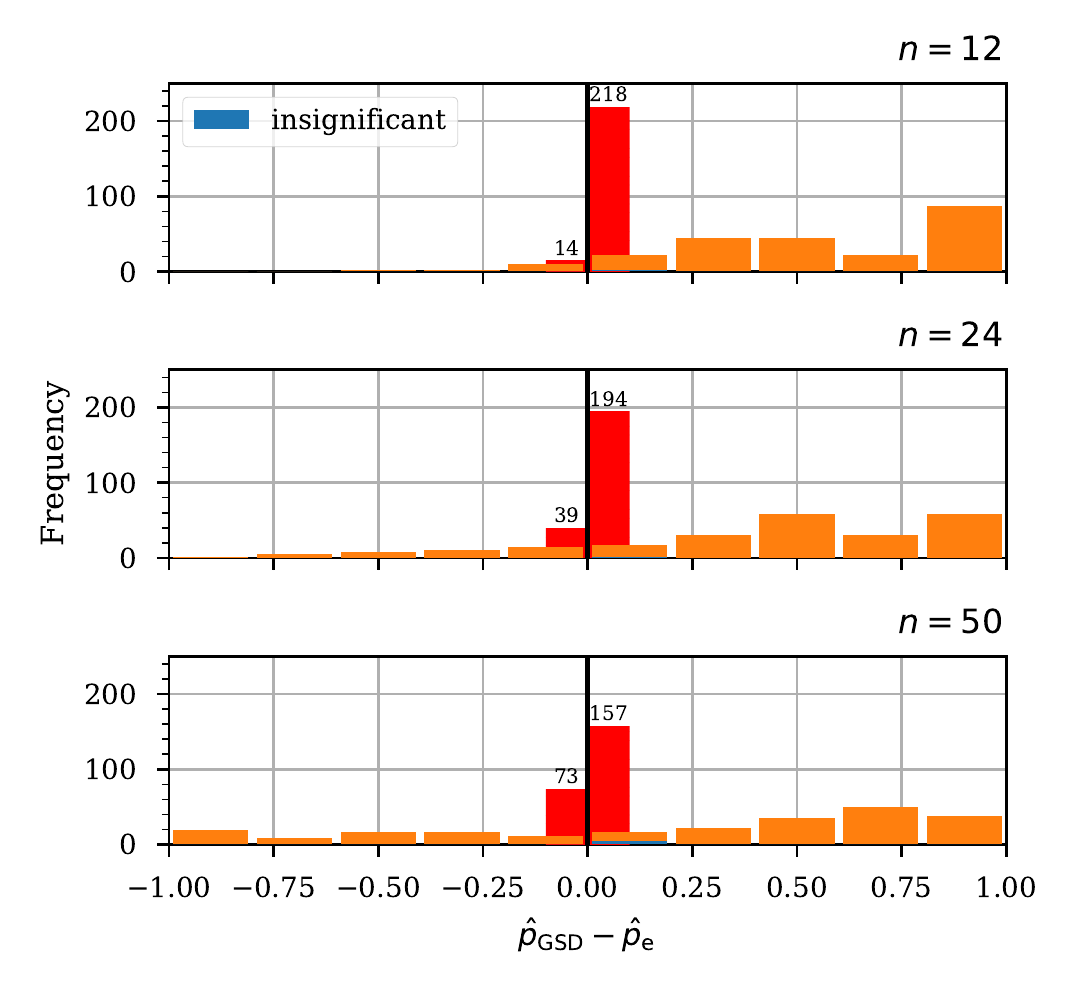}}%
        \label{fig:prob_diff_histograms_corrected}}
    \hfil
    \subfloat[]{\resizebox{.45\textwidth}{!}{\includegraphics[viewport=10 12 300 278, clip]{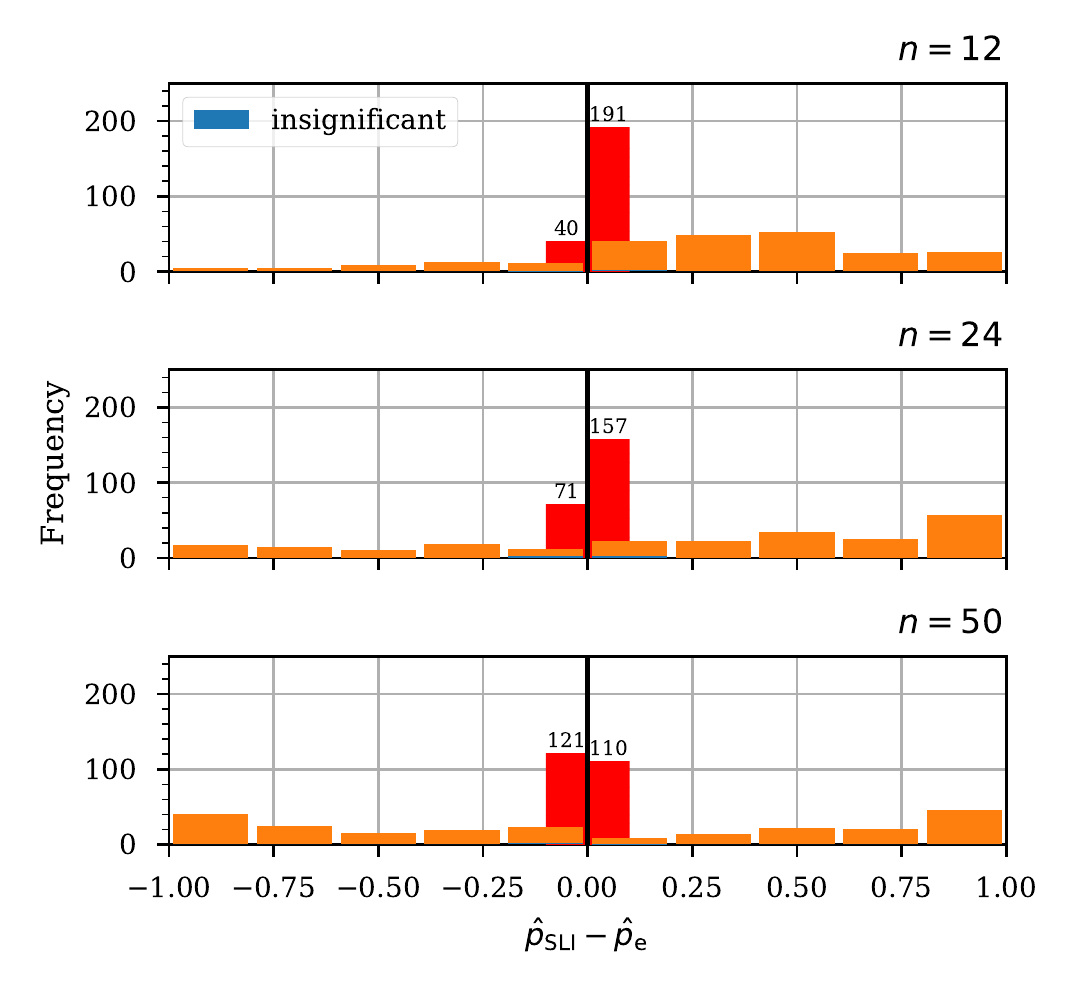}}%
        \label{fig:prob_diff_hist_corr_sli}}
    \caption{Histograms depicting the distribution of probability differences
	$\hat{p}_{\mathrm{GSD}}-\hat{p}_{\mathrm{e}}$ in (a) and
	$\hat{p}_{\mathrm{SLI}}-\hat{p}_{\mathrm{e}}$ in (b).
	Three small	sample sizes are considered: 12, 24 and 50. Blue-coloured
	parts of the bars represent statistically insignificant
	probability differences. Red bars with the hatching
	indicate the sum of probability differences to the right and
	to the left of zero (excluding insignificant results).}
    \label{fig:prob_diff_histograms}
\end{figure*}

Fig.~\ref{fig:prob_diff_histograms_corrected} presents results
of the analysis. They take the form of three histograms.
These histograms visualise probability differences
$\hat{p}_{\mathrm{GSD}}-\hat{p}_{\mathrm{e}}$ for the three
investigated small sample sizes (i.e., 12, 24 and 50).
Now, greater probability mass
to the right of 0 indicates that the GSD performs better than the
empirical distribution. Greater probability mass to the left of 0
corresponds to the opposite situation, i.e., empirical distribution
outperforms the GSD. To make the analysis easier, we show in the plot
red hatched bars that indicate for how many stimuli the GSD outperforms
the empirical distribution (the red hatched bar on the right) and
for how many the empirical distribution performs better than the GSD (the
red hatched bar on the left).
Blue-coloured parts of the
bars represent statistically insignificant probability differences.
When assessing which approach performs better,
these blue parts of the bars are discarded.

Clearly, the GSD outperforms empirical distribution for all three small sample
sizes. The effect is most clearly visible for the small sample size of 12.
As expected, as the size of a small sample grows, empirical distribution's
performance improves. Nevertheless, even for as many as 50 observations
per small sample (which rarely happens in typical MQA subjective
experiments), the GSD still significantly outperforms the empirical
distribution.

Results show
that the GSD is a better choice than the empirical distribution
(which is typically used in this context)
when it comes to resampling of subjective responses from MQA studies. 
This opens up an opportunity to train
better ML models for the MQA applications, without having to
organise large subjective experiments (i.e., experiments with
a large number of participants). This result is yet another
indication of GSD's superiority over methods typically used
for MQA data analysis.

\subsection{Comparison With the State-of-the-Art Model}
\label{ssec:comparison_w_sota_model}
To evaluate GSD's performance against a state-of-the-art solution,
we compare it with the fascinating model presented in~\cite{Li2020EI}.
To make the description easier to comprehend, we refer to the model
from \cite{Li2020EI} as Li2020 model. We selected the Li2020 model
for the comparison, since as of the time of writing this, this is the most
recent and, at the same time, the most popular model in the MQA community.
It should be sufficient to say that the Li2020 model is undergoing
the standardisation process.

The GSD operates only on responses given to a single stimulus.
In other words, the GSD needs to know about only these subject responses that were assigned
to a single stimulus of interest. The Li2020 model requires information regarding all
responses of all subjects that scored the stimulus of interest. Differently put, even though
we are interested in responses assigned to a single stimulus, to estimate model parameters,
we need to know about all responses assigned by a given subject to all other stimuli in the
experiment. Neither the bootstrapped G-test, nor the bootstrapping effectiveness test we use,
satisfy Li2020 model’s requirements. Both tests rely on the assumption that responses
assigned to the single stimulus of interest are sufficient for the model.

Not to abandon the comparison between the GSD and Li2020 models completely, we simplify the
Li2020 model. Specifically, we make it function in a similar manner to the GSD.
Put differently, we make the Li2020 model only require responses assigned to a single
stimulus of interest. This results in a model defined by a Gaussian probability density
function (PDF) with its mean ($\mu$) equal to sample mean (i.e., the MOS) and variance
($\sigma^2$) equal to the sample variance ($s^2$).
(Note that “sample” means here a set of responses
assigned to a single stimulus of interest.) In the following text, we refer to the modified
Li2020 model as the \textit{Simplified Li2020} model or SLI for short.

After estimating Simplified Li2020 model's parameters, we end up with a continuous
normal distribution $\mathcal{N}(\mathrm{MOS}, s^2)$.
Since real subjective responses take the form of discrete
numbers ($\{1, 2, 3, 4, 5\}$ in our case), we need to map from the continuous domain
of the normal distribution to the 5-level scale of interest. For this, we proceed in
the same manner as we do when fitting the Ordered Probit model to the data.
Specifically,
we apply equations (\ref{eq:discretise_234}), (\ref{eq:discretise_1}) and
(\ref{eq:discretise_5}).

Although the Ordered Probit and Simplified Li2020 models look very similar, they
are not identical. The key difference lies in model parameters estimation. The Simplified
Li2020 model assumes observed subjective responses are realisations of a continuous
random variable following the normal distribution. Importantly, realisations
of such a random variable can take any value (from plus to minus infinity). Hence,
observing values from the $\{1, 2, 3, 4, 5\}$ set exclusively is, probabilistically
speaking, very rare. The Simplified Li2020 model ignores this fact and fits the
normal distribution to these data using sample mean and
variance.\footnote{This approach exemplifies what Liddell and Kruschke warn
against in~\cite{Liddell2018}.}
Contrary to the Simplified Li2020 model, the Ordered Probit
model does not assume observed subjective responses are realisations of a continuous
random variable. More precisely, the continuous normal distribution present in
the Ordered Probit model is treated as a latent trait of the data. This latent
continuous distribution is always mapped to a discrete scale of interest first
(cf. (\ref{eq:discretise_234}), (\ref{eq:discretise_1}) and
(\ref{eq:discretise_5})), before fitting the model. Finally, although we
describe here the Simplified Li2020 model, the same discussion applies to
the original Li2020 model. Differently put, the original full model also
assumes that observed subjective responses are realisations of a continuous normal
random variable (even though these responses only take values from the
$\{1, 2, 3, 4, 5\}$ set).

\subsubsection{G-test of Goodness-of-Fit}
\label{sssec:g_test_of_goodness_of_fit}
Let us first check how the SLI model performs when it comes to
describing response distributions observed in typical MQA experiments.
For this, we will use the same G-test-based procedure, as the one we
applied to the GSD in
Sec.~\ref{ssec:good_description_of_typical_mqa_experiments}.
Fig.~\ref{fig:pvalue_pp_plot_mqa_exp} shows the GSD, SLI and Ordered
Probit compared in terms of G-test results. Since only GSD data
points fall below the black diagonal line, it is the only model that
properly reflects response distributions observed in typical MQA
experiments. What is more, the SLI model performs worse both from the
GSD and from the Ordered Probit models. Performance inferior to the
Ordered Probit model may be ascribed to SLI's lack of mapping to
the 5-level scale, when estimating its parameters. In short,
the SLI and Ordered Probit models do not properly describe response
distributions observed in typical MQA experiments, whereas the GSD
model does.

\subsubsection{Bootstrapping}
\label{sssec:bootstrapping}
We now test whether the SLI model can beat the GSD if it comes to
bootstrapping. For this, we apply the same procedure to the SLI model,
as the one we applied to the GSD model in
Sec.~\ref{ssec:better_than_empirical_distribution}. The result
is given in Fig.~\ref{fig:prob_diff_hist_corr_sli}. As we can see,
the SLI model performs better than the empirical probability mass
function (EPMF) for small samples of size 12 and 24. However, it
performs worse than the EPMF for small samples of size 50. When we
compare the results of the SLI, with those of the GSD, we see that
the latter outperforms the former in all cases.


\section{Discussion}
\label{sec:discussion}
Section~\ref{ssec:interpretable_parameters} shows that GSD's parameterisation is more
interpretable and intuitive when compared to Ordered Probit's one.
Importantly, Ordered Probit's parameterisation is not erroneous. Still, using it may
lead to mistaken conclusions if used carelessly. Our results indicate that GSD's
parameterisation should be preferred over Ordered Probit's one. This insight is
relevant for the MQA research community since many practitioners decide to first
try using continuous models (Ordered Probit being one of them) when they start
working with subjective responses modelling. Arguably, their preference to choose
continuous models comes from easier availability of methods operating on
such models. We can also argue that continuous models get more attention during
standard statistics and probability classes and thus naturally come to mind
when thinking about data modelling. We want to protect MQA practitioners against
potential mistakes arising from using continuous models to analyse discrete data.
Our results indicate that discrete models, and the GSD in particular, are viable
and better alternatives to continuous models when it comes to subjective responses
analysis (with responses expressed on a discrete scale).

Section~\ref{ssec:good_description_of_typical_mqa_experiments} reveals that
the GSD well describes responses from typical MQA experiments. This property
of the GSD indicates that the GSD can serve as a basis for building parametric methods
for subjective responses analysis. Please note that parametric methods have
greater power than nonparametric methods do. For example, a parametric hypothesis
testing framework can detect a smaller effect size for a given sample size, when compared
to a nonparametric framework. This increased power may prove essential when analysing
responses from controlled subjective experiments. Since such experiments usually take
place in a laboratory environment and require a direct involvement of a researcher, they
can become resource intensive (both money- and time-wise). It is desirable (or sometimes
necessary) to reduce sample size of such experiments.\footnote{In MQA experiments sample
size usually corresponds to the number of people invited to assess quality of a set of
stimuli.} A parametric GSD-based data analysis framework would help address this problem.
Due to its parametric nature it would be able to detect smaller differences between
various test conditions for a given sample size, in comparison to other nonparametric
methods.

Importantly, neither the GSD nor the Ordered Probit model properly describe response
distributions observed in broadly understood or non-MQA subjective experiments
(cf. Fig.~\ref{fig:p_value_histogram_broad_mqa_exp} and
Fig.~\ref{fig:p_value_histogram_non_mqa_exp}). This
means the models are not globally applicable to modelling subjective responses
expressed on the 5-level Likert scale. Potentially, more complicated models
(i.e., models with more than two parameters) are
necessary to model phenomena present in responses from broadly understood or non-MQA
experiments.

Sec.~\ref{ssec:better_than_empirical_distribution} makes it clear that the
GSD outperforms the traditional approach (based on empirical distribution) when it comes
to subjective responses bootstrapping. This result means that whenever there is a need
to generate more results from a small real-life sample, the GSD should be preferred
over empirical distribution to perform resampling. Such resampling may prove
necessary when building an ML-based perceptual quality predictor. Building such a
predictor requires a significant amount of data. Sample sizes sufficiently large
may be difficult to generate through a controlled experiment. Thus, a small
real-life sample can be collected through a controlled experiment. Then,
the GSD-based bootstrapping can be used to enlarge the small real-life sample
to a larger sample (of a size sufficient for building an ML-based perceptual quality
predictor). Significantly, having such a mechanism at hand also addresses the issue of
controlled experiments being money- and time-intensive. As previously, a small and not so
expensive experiment may be organised to generate a small real-life sample of responses.
This sample can then be enlarged using the GSD-based bootstrapping to achieve sample
size that would otherwise require organising a larger and more expensive controlled
experiment.
We would like to remind the reader at this point that our results indicate
that the mechanism described above applies to responses from typical MQA experiments
exclusively.

Finally, Sec.~\ref{ssec:comparison_w_sota_model} evidences that
the GSD outperforms the state-of-the-art model, namely the Simplified
Li2020 (SLI) model from~\cite{Li2020EI}. GSD's superiority is clear
when it comes to goodness-of-fit testing for data from typical
MQA experiments. Out of the three models tested (GSD, SLI and
Ordered Probit), only the GSD properly describes response
distributions observed in the data. If it comes to bootstrapping,
the SLI, similarly to the GSD, outperforms the empirical
distribution for small sample size of 12 and 24. However, GSD's
improvement over the empirical distribution is larger for the two
cases. Furthermore, only the GSD outperforms the empirical
distribution for the small sample size of 50.


\section{Conclusion}
\label{sec:conclusion}
Our work substantiates the following four claims:
\begin{enumerate}
    \item The GSD has interpretable parameters that clearly and intuitively describe
    response distribution shape (for responses gathered in MQA subjective experiments).
    \item The GSD properly models response distributions observed in typical MQA
    subjective experiments.
    \item The GSD is better suited for bootstrapping of responses from MQA subjective
    experiments in comparison to the traditional approach based on empirical distribution.
    \item The GSD outperforms the state-of-the-art model in terms of
    goodness-of-fit testing (on data from typical MQA experiments)
    and bootstrapping.
\end{enumerate}

The results indicate that the GSD-based bootstrapping of subjective responses from
MQA experiments can be used to build new ML-based perceptual quality predictors,
without having to organise large-scale controlled experiments. This gives a chance
of building ML-based predictors cheaper than would be otherwise possible.

We hope that our discussion regarding interpretable GSD parameters and risks
inherent to using continuous models to analyse discrete subjective responses,
will convince the MQA research community to reconsider current best practices
and recommendations.

There are two directions our future work may take. First, we would like to build
a ML-based perceptual quality predictor. For this we plan to use the GSD-based
bootstrapping. Second, we would like to propose a GSD-based parametric hypothesis
testing framework for analysis of subjective responses from MQA experiments.

At last, we invite everyone to use the GSD to analyse subjective responses
from their experiments and to make use of the tools presented in this paper.
Our GitHub repository (\url{https://github.com/Qub3k/subjective-exp-consistency-check})
contains software tools that make it easier to start using the GSD.
We hope the model and related tools will allow other MQA researchers and
practitioners to analyse their data more efficiently and effectively.


\section*{Acknowledgments}
\label{sec:ack}
\addcontentsline{toc}{section}{Acknowledgments}

We would like to thank Netflix, Inc. for sponsoring initial
stages of this research and for funding they provided to
organise and conduct the NFLX experiment
(cf. Sec.~\ref{ssec:data_sets}). We would also like to thank
Anush Krishna Moorthy for coordinating the NFLX experiment and
providing valuable feedback on a draft version of this work.
We warmly acknowledge comments and suggestions from Zhi Li of
Netflix as well.
\enlargethispage{-0.8in} 


\bibliographystyle{IEEEtran}
\bibliography{bibliography}

\newpage

\appendices

\section{Complete GSD Formulation}
\label{app:full_gsd}
Assuming that $U$ is a discrete random variable describing responses of
a single observer assessing a single stimulus, we can define the GSD as
follows:
\begin{equation}
    U \sim GSD(\psi, \rho), \label{eq:gsd_general}
\end{equation}
where $\psi$ can be understood as a mean opinion of the complete population
of observers and $\rho$ can be understood as a measure of opinion spread.
Now, to reveal GSD's internal structure, we rewrite (\ref{eq:gsd_general}) as follows:
\begin{equation*}
    U \sim \psi + \epsilon,
\end{equation*}
where $\psi$ represents the mean opinion mentioned before
and $\epsilon$ represents an error term
distributed according to the $H_{\rho}$ distribution. Importantly, $H_{\rho}$
satisfies the following five requirements:
\begin{enumerate}
    \item mean equals zero,
    \item variance is linearly dependent on $\rho$,
    \item variance is a decreasing function of $\rho$,
    \item $\rho$ is the only parameter influencing distribution shape (with
    $\rho \in [0, 1]$) and
    \item $H_{\rho}$ models the complete range of possible variances for a
    discrete process with a limited support ($\{1, 2, 3, 4, 5\}$ in our case).
\end{enumerate}

Before we introduce the exact formulation of $H_{\rho}$, let us point out
that the range of possible variances for a discrete process supported
on a $\{1, 2, 3, 4, 5\}$ set is limited. Furthermore, since the mean of
$H_{\rho}$ is zero, the range of variances depends on $\psi$. If we denote
by $V_\mathrm{min}(\psi)$, $V_\mathrm{max}(\psi)$ the minimal and maximal
possible variance, respectively, then
\begin{equation*}
    \begin{split}
    V_{\mathrm{min}}(\psi)&=(\lceil\psi\rceil-\psi)(\psi-\lfloor\psi\rfloor),\\
    V_{\mathrm{max}}(\psi)&=(\psi-1)(5-\psi),
    \end{split}
\end{equation*}
and the interval $[V_{\mathrm{min}}(\psi),V_{\mathrm{max}}(\psi)]$ is the
range of all possible variances. In other words, variance
$\mathbb{V}_{H_{\rho}}(\epsilon)$ of $H_{\rho}$ falls within the
$[V_{\mathrm{min}}(\psi),V_{\mathrm{max}}(\psi)]$ interval.

Let us start by saying that a special case of $H_{\rho}$ is a shifted Binomial
distribution:
\begin{equation*}
    P(\epsilon=k-\psi)=\binom{4}{k-1}\left(\frac{\psi-1}{4}\right)^{k-1}\left(\frac{5-\psi}{4}\right)^{5-k},
\end{equation*}
where $k\in\{1,...,5\}$ represents five response categories. Notice that variance
of this shifted Binomial distribution is equal to:
\begin{equation} \label{eq:bin_var}
    V_{\mathrm{Bin}}(\psi):=\frac{V_{\mathrm{max}}(\psi)}{4}.
\end{equation}
Taking into account that variance of $H_{\rho}$ must linearly depend on $\rho$
and that it must be a decreasing function of $\rho$, we can write:
\begin{equation} \label{eq:h_rho_var}
    \mathbb{V}_{H_{\rho}}(\epsilon)=\rho V_{\mathrm{min}}(\psi)+(1-\rho)V_{\mathrm{max}}(\psi).
\end{equation}
If we now combine (\ref{eq:bin_var}) and (\ref{eq:h_rho_var}), we can come up
with a formula for $\rho$ (for the special case of $H_{\rho}$ when its variance
equals variance of a shifted Binomial distribution):
\begin{equation*}
    \begin{split}
    \mathbb{V}_{H_{\rho}}(\epsilon)=V_{\mathrm{Bin}}(\psi) \  \Rightarrow  \ \rho=C(\psi)\\
    C(\psi):=\frac{3}{4}\ \frac{V_{\mathrm{max}}(\psi)}{V_{\mathrm{max}}(\psi)-V_{\mathrm{min}}(\psi)}.
    \end{split}
\end{equation*}

We can now use variance of a shifted Binomial distribution as a cut-off point
dividing the range of possible variances into two distinct intervals:
\begin{equation*}
    \begin{split}
    \mathbb{V}_{H_{\rho}}(\epsilon)&\in [V_{\mathrm{min}}(\psi),V_{\mathrm{Bin}}(\psi)] \Leftrightarrow \rho \in [C(\psi),1],\\
    \mathbb{V}_{H_{\rho}}(\epsilon)&\in [V_{\mathrm{Bin}}(\psi),V_{\mathrm{max}}(\psi)] \Leftrightarrow \rho \in [0,C(\psi)].
    \end{split}
\end{equation*}

Finally, we can put forward the complete formulation of $H_{\rho}$. For variances bigger
than $V_{\mathrm{Bin}}(\psi)$, $H_{\rho}$ takes the form of a reparameterised
Beta Binomial distribution and is denoted by $G_{\rho}$:
\begin{equation*} \label{eq:Prho}
    \begin{split}
    P_{G_{\rho}}(\epsilon=k-\psi)&= \binom{4}{k-1} \times \\
        &\prod\limits_{i=0}^{k-2}\left(\frac{(\psi-1)\rho}{4}+i(C(\psi)-\rho)\right) \times \\
    	&\frac{\prod\limits_{j=0}^{4-k}\left(\frac{(5-\psi)\rho}{4}+j(C(\psi)-\rho)\right)}{\prod\limits_{i=0}^{3}\left(\rho+i(C(\psi)-\rho)\right)},  
    \end{split}
\end{equation*}
where $\rho\in[0,C(\psi))$ and $k\in\{1,...,5\}$. Notice that for $\rho \rightarrow 0$,
$G_{\rho}$ goes to a two point distribution supported on $\{1-\psi,5-\psi\}$,
with the biggest possible variance equal to $V_{\mathrm{max}}(\psi)$. On the other hand,
for $\rho \rightarrow C(\psi)$, $G_{\rho}$ goes to the shifted Binomial distribution
with variance equal to $V_{\mathrm{Bin}}(\psi)$.

For variances smaller than $V_{\mathrm{Bin}}(\psi)$, $H_{\rho}$ takes the form
of a mixture of two distributions: (i) shifted Binomial distribution and (ii)
two- or one-point distribution, depending on $\psi$ (this corresponds to the distribution
with the smallest possible variance). To make $H_{\rho}$ follow previously stated
requirements, we reparameterise the mixture parameter to make it fit the
$[C(\psi),1]$ interval. We denote the resulting distribution as $F_{\rho}$:
\begin{equation*} \label{eq:Frho}
    \begin{split}
    P_{F_{\rho}} & (\epsilon = k-\psi)= \\ 
    & \frac{\rho-C(\psi)}{1-C(\psi)} [1-|k-\psi|]_{+} + \\
    & \frac{1-\rho}{1-C(\psi)}\binom{4}{k-1}\left(\frac{\psi-1}{4}\right)^{k-1}\left(\frac{5-\psi}{4}\right)^{5-k}, \\
    \end{split} 
\end{equation*}
where $\rho\in[C(\psi),1]$, $[x]_{+} = \max(x,0)$ and $k\in\{1,...,5\}$. Notice that
for $\rho \rightarrow C(\psi)$, $F_{\rho}$ goes to the shifted Binomial distribution
and for $\rho \rightarrow 1$, $F_{\rho}$ goes to a two- or one-point distribution
(depending on $\psi$), with the smallest possible variance $V_{\mathrm{min}}(\psi)$.

At last, we obtain a formula for $H_{\rho}$ and thus for the GSD:
\begin{equation}
    H_{\rho}=G_{\rho}\ I(\rho<C(\psi)) + F_{\rho}\ I(\rho\geq C(\psi)),\tag{GSD}
\end{equation}
where $\psi$ is the expected value $E(U)$, $\rho\in[0,1]$ is a confidence parameter
linearly dependent on variance $V(U)$, i.e.
\begin{equation*}
    \rho=\frac{V_{\mathrm{max}}(\psi)-V(U)}{V_{\mathrm{max}}(\psi)-V_{\mathrm{min}}(\psi)},
\end{equation*}
and $I(x)$ is one if $x$ is true or 0 if $x$ is false.


\section{G-Test, Bootstrap Procedure}
\label{app:gtest}
Denote by $(n_1, n_2, n_3, n_4, n_5)$ numbers of observed responses, i.e. $n_k$
is a number of answers $k$ and $\sum\limits_{k=1}^{5} n_k=n$.
By $(p_1, p_2, p_3, p_4, p_5)$ denote unknown probabilities of the responses $1,\ldots,5$.
We want to test 
$$H_0:\ (p_1, p_2, p_3, p_4, p_5)\ \textrm{are from the GSD}$$ 
against 
$$H_1:\ (p_1, p_2, p_3, p_4, p_5)\ \textrm{are not from the GSD.}$$
One should not use the chi-squared test in case of small numbers in selected cells,
i.e. small $n_k$ for some $k\in\{1,\ldots,5\}$. We use a bootstrap version of
the standard likelihood-ratio test, i.e. G-Test. The procedure is as follows:
\begin{enumerate} 
\item Estimate probabilities of the responses
$(\hat{p}_1, \hat{p}_2, \hat{p}_3, \hat{p}_4, \hat{p}_5)$ using maximum likelihood GSD estimator.
\item Calculate test statistic $T= \sum\limits_{k=1}^{5} n_k \log(n_k / (n \hat{p}_k))$,
where $0 \log(0 / (n \hat{p}_k))=0$.
\item Generate $MC$ (for example $MC=10\,000$) bootstrap samples of size $n$
from the distribution $(\hat{p}_1, \hat{p}_2, \hat{p}_3, \hat{p}_4, \hat{p}_5)$.
Obtain $(m^r_1, m^r_2, m^r_3, m^r_4, m^r_5)$, $r=1,\ldots,MC$, where $m^r_k$ is the
number of responses $k$ in the $r$th bootstrap sample.
\item Estimate probabilities of the responses
$(\hat{q}^r_1, \hat{q}^r_2, \hat{q}^r_3, \hat{q}^r_4, \hat{q}^r_5)$ for every
bootstrap sample $(m^r_1, m^r_2, m^r_3, m^r_4, m^r_5)$ using maximum likelihood GSD estimator. 
\item Calculate bootstrap statistics
\begin{equation*}
T_r= \sum\limits_{k=1}^{5} m^r_k \log(m^r_k / (n \hat{q}^r_k)),
\end{equation*}
where $0 \log(0 / (n \hat{q}^r_k))=0$. 
\item Calculate bootstrap $p$-value using the following equation
\begin{equation*}
p=\frac{1}{MC}\sum\limits_{r=1}^{MC} I(T_r\geq T),
\end{equation*}
where $I(x)$ is one if $x$ is true or 0 if $x$ is false.
\end{enumerate}
Naturally, the procedure
can be applied to models other than the GSD. One has to replace the GSD estimator
in steps 1) and 4), with an estimator appropriate for a model of interest.

\section{Bootstrapping Effectiveness Test}
\label{app:generalisability_test}
This test indicates whether a given model better reflects the distribution
shape of a large sample, when fitted to a subsample of this large sample.
More specifically, the test checks how the model performs when compared
to the empirical probability mass function (EPMF). Importantly, the procedure below
assumes that we are operating on observations expressed on the 5-level
ordinal scale (e.g., 5-level Likert scale). In the field of MQA
this usually corresponds to the following five response categories: 1---Bad,
2---Poor, 3---Fair, 4---Good and 5---Excellent.

Let us denote by $N$ the number of observations in the large sample 
(e.g., $N = 200$) and
by $n$ the number of observations in the subsample of this large sample
(e.g., $n = 24$).
Now, we denote by $\left( N_1, N_2, N_3, N_4, N_5 \right)$ the
frequencies of each response category in the large sample. We denote
by $\left( p_1, p_2, p_3, p_4, p_5 \right)$ the EPMF
of the large sample. The test procedure is as follows:
\begin{enumerate}
    \item Generate $MC$ bootstrap samples (e.g., $MC = 10\,000$) of size $n$
    from the EPMF of the large
    sample $\left( p_1, p_2, p_3, p_4, p_5 \right)$.
    \item For the $r$-th bootstrap sample
    ($r = 1, 2, \ldots, MC$):
    \begin{enumerate}
        \item Estimate response category
        probabilities using maximum likelihood estimation for the model
        of interest (e.g., the GSD model). Denote the estimated
        probabilities by $\left( \hat{q}_1, \hat{q}_2, \hat{q}_3, \hat{q_4},
        \hat{q}_5 \right)$.
        \item Denote by $\left( \hat{v}_1, \hat{v}_2, \hat{v}_3, \hat{v}_4,
        \hat{v}_5 \right)$ the
        EPMF of the bootstrap sample.
        \item Find the likelihood $\mathcal{L}_m$ of the estimated model
        for the large sample.
        In other words, calculate 
        $$
        \mathcal{L}_m = \prod\limits_{k=1, N_k \neq 0}^{5} \hat{q}_k^{N_k}.
        $$
        \item Find the likelihood $\mathcal{L}_e$ of the
        bootstrap sample's EPMF
        for the large sample. In other words, calculate
        $$
        \mathcal{L}_e = \prod\limits_{k=1, N_k \neq 0}^{5} \hat{v}_k^{N_k}.
        $$ 
        \item Find the natural logarithm of the ratio of the two likelihoods
        and denote it by $W_r$
        $$
        W_r = \ln \left( \frac{\mathcal{L}_m}{\mathcal{L}_e}  \right).
        $$
        Note that the above simplifies to
        $$
        W_r = \sum\limits_{k=1, N_k \neq 0}^{5} N_k \left( \ln \hat{q}_k - \ln
        \hat{v}_k \right).
        $$
        \item\label{exeption} (Applies only when the GSD is compared
        to the EPMF.) Check if any of the two exceptions to the point above apply.
        \begin{enumerate}
            \item If the bootstrap sample consists of observations in only
            one response category or in only two neighbouring response categories,
            we know for sure that the performance of the model and the empirical
            distribution is the same and so we set $W_r = 0$ and move with
            the analysis to the subsequent bootstrap sample.
            \item If the condition above is not met and for any response category $k$
            $\hat{v}_k = 0$ and, at the same time, $N_k \neq 0$ then the model-based
            likelihood $\mathcal{L}_m \neq 0$ and the EPMF-based
            likelihood $\mathcal{L}_e = 0$. In such a situation $W_r = \infty$.
            When implementing the test as a software programme we advise to assume
            that $W_r = 10^{10}$.
        \end{enumerate}
    \end{enumerate}
    \item Calculate the estimator of $p_{\mathrm{m}}-p_{\mathrm{e}}=P(W_r > 0)-P(W_r < 0)$, which is the difference between the probability that the model has greater likelihood than the
    EPMF and the probability that the EPMF has greater likelihood than the model.
    This can be formally described by the following.
    $$
    \hat{p}_{\mathrm{m}}-\hat{p}_{\mathrm{e}}= \frac{\sum\limits_{r=1}^{MC} I \left( W_r > 0 \right)}{MC}-\frac{\sum\limits_{r=1}^{MC}I \left( W_r < 0 \right)}{MC},
    $$
    where $I(x)$ is one if $x$ is true or 0 if $x$ is false.
    \item Calculate $.95$ confidence interval for $p_{\mathrm{m}}-p_{\mathrm{e}}$ i.e.,
    $$L=\hat{p}_{\mathrm{m}}-\hat{p}_{\mathrm{e}}-1.96\sqrt{\frac{\hat{p}_{\mathrm{m}}+\hat{p}_{\mathrm{e}}-(\hat{p}_{\mathrm{m}}-\hat{p}_{\mathrm{e}})^2}{MC}}$$
    $$R=\hat{p}_{\mathrm{m}}-\hat{p}_{\mathrm{e}}+1.96\sqrt{\frac{\hat{p}_{\mathrm{m}}+\hat{p}_{\mathrm{e}}-(\hat{p}_{\mathrm{m}}-\hat{p}_{\mathrm{e}})^2}{MC}}$$
    For $L>0$ the model performs better. For $R<0$ the EPMF performs better.
    If $[L,R]$ contains zero there is no significant difference between
    the model and EPMF.
    
\end{enumerate}
\subsection{Parameters Estimation Modification}
\label{app:parameters_estimation_modification}
In our analyses, we apply the above procedure using a modified probability estimators for
the GSD, SLI and empirical distribution. We need such a modification, since if for a specific item we collected $n$ answers, and none of them is category $i$, it does not mean that in a larger sample category $i$ cannot be observed. It only means that with the number of answers under consideration, $n$ in our case, the specific category, $i$ in our example, was not observed. Therefore, concluding that the probability of observing category $i$ is $0$ should be modified to a positive value, smaller for larger $n$. Let us denote this value by $\epsilon(n)$.
Significantly, one of the consequences of such modification is preventing any response category probability from being equal to $0$.

It is an open question how to define $\epsilon(n)$. For the empirical distribution, we can simply add $0.5$ to all response category counts (cf. \cite{pagano2018principles}), i.e.,
$$
\forall k \in \{ 1, \cdots, 5 \}, \quad \hat{v}_k=\frac{n_k+0.5}{n+2.5},
$$
where $n_k$ is the response count of category $k$ in a bootstrap sample.

For the GSD it is enough to estimate parameters $\psi,\rho$ on the set $[1+\epsilon_{\psi_d}(n),5-\epsilon_{\psi_u}(n)]\times[0+\epsilon_{\rho_{\mathrm{d}}}(n),1-\epsilon_{\rho_{\mathrm{u}}}(n)]$, where $\epsilon_{\psi_{\cdot}}(n)>0$, $\epsilon_{\rho_{\cdot}}(n)>0$ and $\lim\limits_{n \rightarrow \infty} \epsilon_{\psi_{\cdot}}(n)=\lim\limits_{n \rightarrow \infty} \epsilon_{\rho_{\cdot}}(n)=0$.
So the problem of defining $\epsilon(n)$ is changed to defining $\epsilon_{\psi_{\cdot}}(n)$ and $\epsilon_{\rho_{\cdot}}(n)$. 

To define $\epsilon_{\psi_{\cdot}}(n)$ and $\epsilon_{\rho_{\cdot}}(n)$ we introduce
a limit for the maximum probability any two response categories can add up to (and call
it $p_{\max}$). Importantly, when assessing $p_{\max}$, we only take into account two
most probable response categories. This can be formally written as follows:
\begin{equation}\label{eq:pmax}
\begin{split}
    p_{\max} &= P(U=j) + P(U=k), \textrm{ where: }\\
    j&: P(U=j) = \max_{i\in[1,5]} P(U=i) \textrm{ and }\\
    k&: P(U=k) = \max_{i\in[1,5], i \neq j} P(U=i).
\end{split}
\end{equation}

The final algorithm for fitting the GSD to a sample is as follows. Find such
($\hat{\psi}$, $\hat{\rho}$) that satisfy the following two criteria:
\begin{enumerate}
    \item $p_{\max} \leq 1 - \frac{1}{n}$, where $p_{\max}$ is given by (\ref{eq:pmax}) and $n$ is the sample size, and
    \item the likelihood function has the maximum value.
\end{enumerate}

\begin{figure}[!t]
	\centering
	\includegraphics[width=0.48\textwidth,clip,viewport=10 0 370 295]{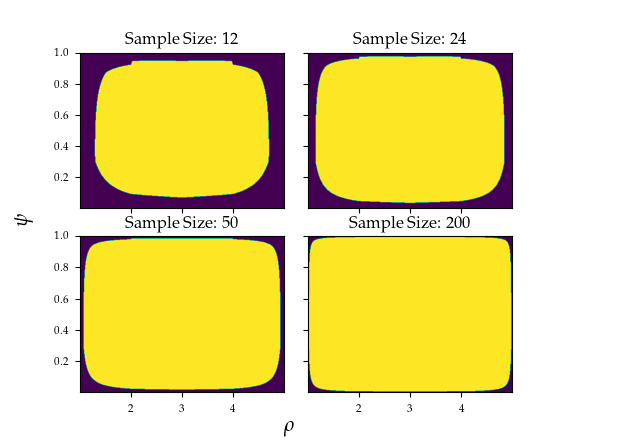}
	\caption{Boundary for $\psi$ and $\rho$ for a given sample size $n$ and $p_{\max}\leq 1 - \frac{1}{n}$. Yellow color marks ($\psi$, $\rho$) pairs considered in the MLE algorithm.}
	\label{fig:psi_rho_boundary}
\end{figure}

An example of $\psi$ and $\rho$ ranges for different sample sizes
is shown in Fig.~\ref{fig:psi_rho_boundary}. 

For the SLI, it is sufficient to introduce a lower limit on $\hat{\sigma}$.
Specifically, if $\hat{\sigma} < c_n$, then $\hat{\sigma} = c_n$, where
$$
c_n = \frac{1}{2 Q(1 - \frac{1}{2n})},
$$
with $Q(p)$ being the quantile function of the standard
normal distribution for probability $p$. This modification makes
sure that the most probable response category gets at most
$1/n$ of the probability mass.
\end{document}